\documentclass[a4paper,11pt]{article}

\usepackage{jheppub_mod} 
\usepackage[T1]{fontenc} 

\usepackage{hyperref}
\usepackage{graphicx}
\usepackage{amsmath,amssymb,slashed}
\usepackage{booktabs,tabulary}
\usepackage{dsfont}
\usepackage{color}
\usepackage[dvipsnames]{xcolor}
\usepackage{multirow}
\usepackage{subcaption}
\usepackage{tikz-feynman}

\newcommand{\Fpi}{F_\pi}
\newcommand{\mpi}{M_{\pi}}

\newcommand{\meta}{M_{\eta}}
\newcommand{\metap}{M_{\eta'}}
\newcommand{\kappapi}{\kappa\pipi}
\newcommand{\kappaeta}{\kappa\etapi}
\newcommand{\Order}{\mathcal{O}}

\newcommand{\GeV}{\,\text{GeV}}
\newcommand{\BR}{\text{BR}}
\newcommand{\beq}{\begin{equation}}
\newcommand{\eeq}{\end{equation}}
\newcommand{\diff}{\text{d}}
\newcommand{\sth}{s_\text{th}}
\newcommand{\tth}{t_\text{th}}
\newcommand{\uth}{u_\text{th}}
\newcommand{\M}{\mathcal{M}}
\newcommand{\F}{\mathcal{F}}
\newcommand{\A}{\mathcal{A}}
\newcommand{\T}{\mathcal{T}}
\newcommand{\G}{\mathcal{G}}
\renewcommand{\H}{\mathcal{H}}

\newcommand{\disc}{\text{disc}\,}
\renewcommand{\Re}{\text{Re}\,}
\renewcommand{\Im}{\text{Im}\,}
\newcommand{\chpt}{$\chi$PT}
\newcommand{\pipi}{_{\pi\pi}}
\newcommand{\etapi}{_{\eta\pi}}
\newcommand{\etap}{\eta^{(\prime)}}

\allowdisplaybreaks[1]

\title{\boldmath{Patterns of $C$- and $CP$-violation in hadronic\\ $\eta$ and $\eta'$ three-body decays}}

\author[a]{Hakan Akdag,}
\author[a,b]{Tobias Isken,}
\author[a]{and Bastian Kubis}

\affiliation[a]{
Helmholtz-Institut f\"ur Strahlen- und Kernphysik (Theorie) and \\
Bethe Center for Theoretical Physics, Universit\"at Bonn, 53115 Bonn, Germany}

\affiliation[b]{Helmholtz Forschungsakademie Hessen f\"ur FAIR (HFHF) and \\
GSI Helmholtzzentrum f\"ur Schwerionenforschung GmbH,\\
Planckstra{\ss}e 1, 64291 Darmstadt, Germany}

\emailAdd{akdag@hiskp.uni-bonn.de}
\emailAdd{isken@hiskp.uni-bonn.de}
\emailAdd{kubis@hiskp.uni-bonn.de}

\abstract{We construct hadronic amplitudes for the three-body decays
  $\eta^{(\prime)}\to\pi^+\pi^-\pi^0$ and $\eta'\to\eta\pi^+\pi^-$ in a non-perturbative fashion,
  allowing for $C$- and $CP$-violating asymmetries in the $\pi^+\pi^-$ distributions.
  These amplitudes are consistent with the constraints of analyticity and unitarity. We find that the currently most accurate Dalitz-plot distributions taken by the KLOE-2 and BESIII collaborations confine the patterns of these asymmetries to a relative per mille level.
  Our dispersive representation 
  allows us to extract the individual coupling strengths of the $C$- and $CP$-violating contributions arising from effective isoscalar and isotensor operators in $\eta^{(\prime)}\to\pi^+\pi^-\pi^0$ and an effective isovector operator in $\eta'\to\eta\pi^+\pi^-$, while the strongly different sensitivities to these operators can be understood from chiral power counting arguments.
}

\begin{document} 
\maketitle

\section{Introduction}
\label{sec:intro}
The idea that the conservation of discrete symmetries in the strong interactions does not have to be manifest first rose to prominence in the year 1950 with the work of Purcell and Ramsey~\cite{Purcell:1950zz}, who proposed the violation of $P$ and $CP$ in the decay~$\eta \to 2 \pi$. Later, this idea was theoretically realized in a $P$- and $CP$-odd operator of dimension four in QCD, which is well known as the $\theta$-term. The latter induces, amongst others, an electric dipole moment (EDM) of the neutron. Rigorous experimental limits on EDMs imply corresponding theoretical limits on $\eta \to 2 \pi$~\cite{Shifman:1979if,Crewther:1979pi,Pich:1991fq}, a link that can even be established without recourse to the $\theta$-term as the fundamental mechanism~\cite{Gorchtein:2008pe,Gutsche:2016jap,Zhevlakov:2018rwo,Zhevlakov:2019ymi,Zhevlakov:2020bvr,Gan:2020aco}. Accordingly, no measurement so far could find evidence for this process, which is probably beyond experimental reach for the foreseeable future.

Given the dearth of experimental evidence for sources of $CP$-violation beyond the Cabibbo--Kobayashi--Maskawa mechanism in the weak interactions of the Standard Model (SM), it is worthwhile to investigate another category of $CP$-violating operators that has gained much less attention so far: $T$-odd and $P$-even (TOPE) interactions, which in addition violate $C$ according to the $CPT$ theorem. 
$C$-violating effective operators have been discussed in the literature to some extent~\cite{Khriplovich:1990ef,Conti:1992xn,Engel:1995vv,Ramsey-Musolf:1999cub,Kurylov:2000ub}, but explicit links to hadronic processes have barely been established.
In the Standard Model effective field theory~\cite{Buchmuller:1985jz,Grzadkowski:2010es}, they only contribute starting at dimension 8~\cite{Li:2020gnx,Murphy:2020rsh}.\footnote{We assume throughout that additional $C$- and $CP$-violation originates from physics at some high-energy scale beyond the electroweak one, and do not discuss the possibility of it being induced by light, extremely weakly coupled particles.}
As TOPE forces cannot be mediated by $\pi^0$ exchange~\cite{Simonius:1975ve}, they can only contribute via short-range nuclear forces, and are therefore far less constrained from nuclear physics. Suitable candidates to investigate these kinds of operators are certain decays of the $\etap$ mesons, which are eigenstates of $C$.  These allow us to investigate TOPE forces in the absence of the weak interaction, such that the observation of a corresponding $C$-violating $\etap$ decay would automatically indicate physics beyond the Standard Model (BSM).
	
Studying the charge asymmetry of the $\eta\to\pi^{+}\pi^{-}\pi^{0}$ Dalitz-plot distribution offers an ideal stage in the search for such BSM physics.  As pointed out in Ref.~\cite{Gardner:2019nid}, in contrast to other $C$-violating processes such as $\etap\to3\gamma$, $\etap\to\pi^0\gamma^*$, etc., the breaking of mirror symmetry in $\eta\to\pi^{+}\pi^{-}\pi^{0}$ is linear in these BSM operators, as it is generated through interference with the SM mechanism. For an overview of $C$- and $CP$-violating processes in the $\eta$ and $\eta'$ sector we suggest Ref.~\cite{Gan:2020aco}.   The simplest observable that can be probed experimentally is the left-right asymmetry $A_{LR}$ that compares the two halves of the Dalitz-plot distribution divided along the $\pi^{+}\!\leftrightarrow\pi^{-}$ line of reflection~\cite{Layter:1972aq}. It is also possible to construct more sophisticated quadrant and sextant asymmetry parameters $A_{Q}$ and  $A_{S}$ that allow us to disentangle the contributions of the BSM $\Delta I=0,2$ operators, respectively~\cite{Lee:1965zza,Layter:1972aq,Nauenberg:1965}. The KLOE-2 collaboration, in the most precise measurement of the $\eta\to\pi^+\pi^-\pi^0$ Dalitz plot to date, reports all three asymmetry parameters to be consistent with zero~\cite{Anastasi:2016cdz}, superseding many earlier experimental investigations~\cite{Gormley:1968zz,Gormley:1970qz,Layter:1972aq,Jane:1974mk,Ambrosino:2008ht,WASA-at-COSY:2014wpf,BESIII:2015fid}.  Alternatively, $C$-violation in the phenomenological expansion of the Dalitz-plot distribution, i.e., a two-dimensional Taylor series around its center, can be studied by allowing for both $C$-conserving and $C$-violating terms. Until now the KLOE-2 collaboration has probed the first four $C$-violating terms of this parameterization, which again are all consistent with zero~\cite{Anastasi:2016cdz}. Thus, experimentally there is no evidence found for $C$-violation in $\eta\to\pi^{+}\pi^{-}\pi^{0}$.

Theoretical studies of $C$-violation in $\eta\to\pi^{+}\pi^{-}\pi^{0}$ first came to prominence~\cite{Lee:1965zza,Layter:1972aq,Nauenberg:1965} after the discovery of $CP$-violating $K_{L}^{0}\to\pi\pi$ decays in the 1960s~\cite{Christenson:1964fg,Lee:1965hi}. Already at this time it was claimed that $\eta\to\pi^{+}\pi^{-}\pi^{0}$ is far more sensitive to isotensor $\Delta I=2$ than to isoscalar $\Delta I=0$ transitions, since the latter is suppressed by a large angular momentum barrier~\cite{Prentki:1965tt}. Effective BSM operators $X^{\slashed{C}}_{I}$ for $\eta\to\pi^{+}\pi^{-}\pi^{0}$ are given by
\begin{equation}
\begin{aligned}
	X_{0}^{\slashed{C}}&\sim\epsilon_{ijk}\,(\partial_{\mu}\partial_{\nu}\partial_{\lambda}\pi^{i})(\partial^{\mu}\partial^{\nu}\pi^{j})(\partial^{\lambda}\pi^{k})\,\eta\,,\\[.2cm]
	X_{2}^{\slashed{C}}&\sim\epsilon_{ij3}\,\pi^{i}\,(\partial_{\mu}\pi^{j})(\partial^{\mu}\pi^{3})\,\eta\,,
	\label{eq:BSM-operators}
\end{aligned}
\end{equation}
involving at least six derivatives for a $\Delta I=0$ transition, while for $\Delta I=2$ only two derivatives are required. 
The above operators imply a strong kinematic suppression of the $\Delta I=0$ transition compared to $\Delta I=2$ across the Dalitz plot, given the small available phase space in $\eta\to\pi^{+}\pi^{-}\pi^{0}$, as long as the respective coupling strengths of both operators are of similar size.

However, since the 1960s $C$-violation in this decay has been mostly neglected by theory until recently a new theoretical formalism was proposed in Ref.~\cite{Gardner:2019nid}. In this framework the decay amplitude is decomposed into three contributions that can be associated with operators describing the isospin transitions $\Delta I=0,1,2$. While the Standard-Model contribution is driven almost exclusively by the $\Delta I=1$ contribution (ignoring isospin breaking of higher order that is known to have only tiny effects~\cite{Ditsche:2008cq,Schneider:2010hs}), the additional BSM amplitudes arise from $\Delta I=0,2$ transitions. The individual strengths of the latter are given by two complex-valued normalizations. Physically this approach is more meaningful compared to simple phenomenological (i.e., polynomial) parameterizations, as it allows for a direct extraction of the coupling strengths that may subsequently be matched to underlying BSM operators.
The energy dependence of the $C$-violating amplitudes in Ref.~\cite{Gardner:2019nid} is based on the well-known one-loop representation of the SM decay in chiral perturbation theory ($\chi$PT)~\cite{Gasser:1984pr}.  
The authors find the BSM normalization of the $\Delta I=0$ amplitude to be between two and four orders of magnitude less rigorously constrained than the $\Delta I=2$ one, which is a result of the predicted kinematic suppression of the $\Delta I=0$ transition~\cite{Prentki:1965tt}, but again there is no hint for $C$-violation in $\eta\to\pi^{+}\pi^{-}\pi^{0}$ as both BSM normalizations are consistent with zero.

A more rigorous construction of the BSM amplitudes consistent with the fundamental principles of analyticity (a mathematical description of causality) and unitarity (a consequence of probability conservation) can be achieved with techniques from dispersion theory, using the so-called Khuri--Treiman representations~\cite{Khuri:1960zz}.   As Sutherland's theorem~\cite{Sutherland:1966zz,Bell:1996mi}, a statement of current algebra, and $\chi$PT calculations~\cite{Baur:1995gc,Ditsche:2008cq} proved that electromagnetic effects are tiny compared to isospin breaking due to the light quark mass difference $m_u-m_d$, modern dispersion-theoretical studies of the SM contribution $\eta\to3\pi$~\cite{Kampf:2011wr,Guo:2015zqa,Guo:2016wsi,Colangelo:2016jmc,Albaladejo:2017hhj,Colangelo:2018jxw,Kampf:2019bkf} 
focus on a consistent, non-perturbative description of the final-state interactions with the goal to provide information on these fundamental SM parameters. Such a treatment of final-state interactions can also be incorporated in the $C$-violating amplitudes by establishing the corresponding dispersion relations for the $\Delta I=0,2$ transitions.
As a by-product, such dispersive amplitude representations allow us to argue more rigorously why the dependence on yet unknown short-distance operators can be subsumed in a single unknown multiplicative constant for each isospin.

The opportunity to investigate $C$-odd effects as an interference in a Dalitz plot exists similarly for the decay $\eta'\to\eta\pi^+\pi^-$ (although without the potential benefit of the SM decay being suppressed by isospin). This is particularly interesting as the possible asymmetry in the distribution of the charged pions in this decay is sensitive to a different class of $C$-violating operators from those constrained in $\eta\to\pi^+\pi^-\pi^0$, namely the ones with $\Delta I=1$.  Both decays therefore provide orthogonal probes as far as the isospin structure of the $C$-violating operators is concerned.  For $\eta'\to\eta\pi^+\pi^-$ such an operator must include two derivatives and explicitly reads
 \begin{equation}
	X_{1}^{\slashed{C}}\sim\epsilon_{ij3}\,\pi^{i}\,(\partial_{\mu}\pi^{j})(\partial^{\mu}\eta)\,\eta'\,.
	\label{eq:BSM-operator etap}
\end{equation}
The experimental limits on the left-right asymmetry $A_{LR}$ and the $C$-odd contributions of the phenomenological Dalitz-plot expansion measured by the BESIII collaboration~\cite{BESIII:2017djm} vanish again within one standard deviation. Prior to that, measurements by VES~\cite{Dorofeev:2006fb} as well as an earlier BESIII result~\cite{BESIII:2010niv} came to the same conclusion, albeit with much lower accuracy.  While the theoretical description of the SM contribution relying on a sophisticated dispersion-theoretical approach was first established in Refs.~\cite{Schneider:2012nng,Isken:2017dkw}, the incorporation of $C$-violating effects is missing so far.

In this work we generalize the dispersion-theoretical analysis of $C$-conserving SM $\eta\to3\pi$ and $\eta'\to\eta\pi\pi$ decays to additional $C$-violating BSM contributions. Accordingly, the presented dispersive representations account for a consistent resummation of the respective three-particle final-state interactions in all allowed isospin transitions. 
To establish dispersion relations for the new $C$-violating contributions we split our analysis into two parts, Sect.~\ref{sec:eta} dealing with $\eta\to3\pi$ and Sect.~\ref{sec:ep-epp} with $\eta'\to\eta\pi\pi$, and follow the same general strategy in both of them. 
We start with the definition of the $T$-matrix elements and the general kinematics of the respective process in Sects.~\ref{sec:kinematics eta} and~\ref{sec:kinematics eta'}. In Sects.~\ref{sec:reconstruction theorem eta} and~\ref{sec:reconstruction theorem eta'} we decompose the amplitudes into ones depending on one Mandelstam variable only, tremendously simplifying the evaluation. These single-variable amplitudes (SVA) are constrained by elastic unitarity as described in Sects.~\ref{sec:elastic unitarity eta} and~\ref{sec:elastic unitarity eta'}. Sections~\ref{sec:TaylorInvariants Eta} and~\ref{sec:TaylorInvariantsEta'} describe how to extract coupling constants for the effective BSM operators in terms of the subtraction constants. The latter are the free parameters of our dispersive representation, which are fixed by a $\chi^{2}$-fit to data in Sects.~\ref{sec:FixingSubcons Eta} and~\ref{sec:FixingSubConsEta'}. Afterwards we compare our representations of the three-body amplitudes to measurements of the corresponding Dalitz-plot distributions and theoretical constraints in Sects.~\ref{sec:Observables Eta} and~\ref{sec:observables eta'}. 
Section~\ref{sec:eta'} contains a brief comment on how to generalize the analysis for $\eta\to\pi^+\pi^-\pi^0$ to $\eta'\to\pi^+\pi^-\pi^0$.
We conclude our study with a summary covering both parts in Sect.~\ref{sec:summary}.

\section{Dispersive representation of $\boldsymbol{\eta\to3\pi}$}
\label{sec:eta}
As our first step towards the analysis of TOPE forces, we will investigate the decay $\eta \to \pi^+ \pi^- \pi^0$. For this purpose we rely on the sophisticated and well-established Khuri--Treiman framework~\cite{Khuri:1960zz}, in which a set of integral equations for the scattering process $\eta\pi\to\pi\pi$ is established. For the corresponding  dispersion relations we only take the dominant elastic pion--pion rescattering into account. With an analytic continuation of the decay mass as well as the Mandelstam variables one can project onto the physical realm of the decay, thereby taking final-state interactions to all orders in perturbation theory into account, and at the same time obtain a manifestly unitary amplitude.

Before going into more detail, let us have a look at the general properties of the $\eta \to \pi^+ \pi^- \pi^0$ amplitude. Regarding the involved quantum numbers, Bose symmetry demands that $C=(-1)^{I+1}$~\cite{Lee:1965zza}, where $I$ is the total isospin of the three-body final state, which has to be distinguished from the isospin of the decaying meson. In analyses consistent with the symmetries of the Standard Model~\cite{Osborn:1970nn,Gasser:1984pr,Kambor:1995yc,Anisovich:1996tx,Bijnens:2002qy,Borasoy:2005du,Bijnens:2007pr,Schneider:2010hs,Kampf:2011wr,Guo:2015zqa,Guo:2016wsi,Colangelo:2016jmc,Albaladejo:2017hhj,Colangelo:2018jxw,Kampf:2019bkf}, the decay amplitude is exclusively driven by isospin-breaking effects. The reason for this lies in the fact that this decay breaks $G$-parity, whose prerequisite is that either isospin or charge conjugation symmetry is broken, or both. As Standard-Model analyses consider the latter the more cherished symmetry (disregarding the weak interactions), the corresponding amplitudes solely contain $\Delta I=1$ transitions.\footnote{The $C$-conserving $\Delta I=3$ transition is strongly suppressed.} On the contrary, in this work we allow for even isospin transitions $\Delta I=0,2$ and hence imply $C$-violation. Moreover, considering that all involved particles are pseudoscalars, the decay at hand preserves parity and one can conclude that $CP$ has to be violated, too. In summary, the $C$-violating mechanisms are driven by isoscalar $\Delta I=0$ or isotensor $\Delta I=2$ operators~\cite{Lee:1965zza,Prentki:1965tt,Nauenberg:1965,Barrett:1965ia}, such that the generalized $\eta\to\pi^{+}\pi^{-}\pi^{0}$ amplitude has to be of the form~\cite{Gardner:2019nid}
\beq\label{eq:FullAmpEta3Pi}
    \M_c(s,t,u)= \M^{\not C}_0(s,t,u) + \xi\,\M^C_1(s,t,u) + \M^{\not C}_2(s,t,u), 
\eeq
which is split into a contribution for each total isospin denoted by the respective index. In accordance with Refs.~\cite{Colangelo:2016jmc,Colangelo:2018jxw}, we factorized out the isospin-breaking normalization of the SM amplitude
\beq
\xi = \frac{\hat M_{K^+}^2-\hat M_{K^0}^2}{3\sqrt{3}F^2_\pi} = -0.140(9)
\eeq
in terms of the pion decay constant $F_\pi$ and the QCD kaon mass difference.  The isoscalar amplitude $\M^{\not C}_0$ is isospin-conserving but $C$-violating, the Standard-Model amplitude $\M^{C}_1$ is isospin-violating but $C$-conserving, and the isotensor contribution $\M^{\not C}_2$ violates both quantum numbers.  Note that isospin symmetry is an accidental (approximate) symmetry of the strong interactions due to the smallness of the two lightest quark masses (as well as their difference) on typical hadronic scales; as we do not know anything about the isospin structure of the BSM operators, there is no reason to assume isospin to be a useful symmetry for them, too, and hence imply any kind of hierarchy between isoscalar and isotensor $C$-violation on the underlying, fundamental level.

As a further consequence of Bose symmetry, the $C$-violating operators can only contribute to the charged decay mode, but not to $\eta\to3\pi^{0}$. The latter is thus solely given in terms of $\M_1^C$ and explicitly reads
\beq
    \M_n(s,t,u)=\xi\,\big[\M_1^C(s,t,u)+\M_1^C(t,u,s)+\M_1^C(u,s,t)\big],
    \label{eq:isospin-relation}
\eeq
as demanded by isospin symmetry.  Corrections to Eq.~\eqref{eq:isospin-relation} arise only due to higher-order corrections such as virtual-photon effects or the charged-to-neutral pion mass difference~\cite{Ditsche:2008cq,Schneider:2010hs}.

As the Standard-Model contribution $\M_1^C$ has already been extensively studied using Khuri--Treiman equations in Refs.~\cite{Kambor:1995yc,Anisovich:1996tx,Kampf:2011wr,Guo:2015zqa,Guo:2016wsi,Colangelo:2016jmc,Albaladejo:2017hhj,Colangelo:2018jxw,Kampf:2019bkf}, in this section we generalize the dispersive analysis by elaborating on the $C$- and $CP$-odd amplitudes $\M_0^{\not C}$ and $\M_2^{\not C}$.

\subsection{Kinematics}
\label{sec:kinematics eta}
Let us define the $\eta \to \pi^+ \pi^- \pi^0$ transition amplitude in the common manner
\beq
    \big\langle \pi^+(p_+)\,\pi^-(p_-)\,\pi^0(p_0)\big|iT\big| \eta(P_\eta)\big\rangle
    =i\,(2\pi)^4\,\delta^{(4)}( P_\eta-p_+-p_--p_0)\,\M_c(s,t,u)\,.
\eeq
Up to the overall isospin-breaking normalization, we work in the isospin limit, i.e., $\mpi\equiv M_{\pi^\pm}=M_{\pi^0}$, and conventionally write the corresponding Mandelstam variables as 
\beq
    s = ( P_\eta-p_0)^2\,, \qquad
    t = ( P_\eta-p_+)^2\,, \qquad
    u = ( P_\eta-p_-)^2\,,  
\eeq
fulfilling the relation 
\beq\label{eq:MandelstamSum}
    s+t+u=M_{\eta}^2+3M_\pi^2\equiv 3r\,.
\eeq
Note that the amplitude $\M^C_1$ is symmetric under $t\leftrightarrow u$, while $\M^{\not C}_0$ and $\M^{\not C}_2$ are both antisymmetric under the exchange of these two Mandelstam variables.
In the two-pion center-of-mass system, $t$ and $u$ can be expressed in terms of $s$ and the $s$-channel scattering angle $z_s$ by
\beq
    t(s,z_s) = u(s,-z_s)=\frac{1}{2}\big( 3r-s+\kappa(s)z_s \big)\,,
\eeq
with 
\beq\label{eq:Kappa}
    z_s=\cos\theta_s=\frac{t-u}{\kappa(s)}\,,\qquad \kappa(s)=\sigma(s)\,\lambda^{1/2}(M_\eta^2,M_\pi^2,s)\,,
\eeq
where $\sigma(s)=\sqrt{1-4M_\pi^2/s}$ and $\lambda(x,y,z)=x^2+y^2+z^2-2(xy+xz+yz)$ denotes the Källén function.

\subsection{Reconstruction theorem}
\label{sec:reconstruction theorem eta}
To avoid the intricate analysis of complex functions depending on multiple variables one can exploit a decomposition of the amplitude into single-variable functions. Such a decomposition is commonly referred to as \textit{reconstruction theorem}. It was first proven that the latter holds exactly up to and including two-loop order in the framework of \chpt\ for $\pi\pi$ scattering~\cite{Stern:1993rg}, followed by generalizations for unequal masses~\cite{Ananthanarayan:2000cp} and scattering of mesons belonging to the pseudoscalar octet~\cite{Zdrahal:2008bd}.

Neglecting the discontinuities of $D$- and higher partial waves, one may express each amplitude of \textit{total} isospin, on the right-hand side of Eq.~\eqref{eq:FullAmpEta3Pi}, in terms of functions depending on only one kinematical variable, the relative angular momentum and isospin of the $\pi\pi$ intermediate state~\cite{Kambor:1995yc,Anisovich:1996tx,Gardner:2019nid}:
\begin{equation}
\begin{aligned}\label{eq:reconstruction theorems}
    \M^C_1(s,t,u)&=\F_0(s)+(s-u)\,\F_1(t)+(s-t)\,\F_1(u)+\F_2(t)+\F_2(u)-\frac{2}{3}\F_2(s)\,,
    \\
    \M_0^{\not C}(s,t,u)&=(t-u)\,\G_1(s) + (u-s)\,\G_1(t)+(s-t)\,\G_1(u)\,,
    \\[.1cm]
	\M_2^{\not C}(s,t,u)&=2(u-t)\,\H_1(s) + (u-s)\,\H_1(t)+(s-t)\,\H_1(u)-\H_2(t)+\H_2(u)\,.
\end{aligned}
\end{equation}
Due to Bose symmetry, the isospin $I$ of the two-pion state fixes the partial wave $\ell$ unambiguously by means of $\A_{0,2}\equiv\A^{\ell=0}_{I=0,2}$ and $\A_1\equiv\A^{\ell=1}_{I=1}$, with $\A\in\{ \F,\G,\H \}$. 
Note that the single-variable functions $\F$, $\G$, and $\H$ are completely decoupled and can be evaluated independently. Furthermore, each single-variable function $\A_I(s)$ has only a right-hand cut. At this point the charge asymmetry in the $C$-odd contributions, stemming from the exchange $t\leftrightarrow u$, becomes evident.
It is worth noting that the decomposition into single-variable functions presented here is not unique. The relation between the Mandelstam variables given in Eq.~\eqref{eq:MandelstamSum} allows us to shift the amplitude by polynomials in $s$, $t$, and $u$, i.e., $\A_I\to\A_I +\Delta \A_I$, without affecting the reconstruction theorems. The five-parameter ambiguity for the Standard Model reads~\cite{Colangelo:2018jxw}
\beq\label{eq:AmbiguityF}
\begin{split}
    \Delta \F_0(s)&=-4a_1+b_1\,(5s-9r)-3c_1\,(s-r)-27d_1\,r\,(s-r)\\[.1cm]
    &\hspace{2.9cm} +4d_1\,s^2-162e_1\,r^2\,(s-r)-4e_1\,s^2\,\,,
    \\[.1cm]
    \Delta \F_1(s)&=c_1+3d_1\,s+9e_1\,s^2\,,
    \\[.1cm]
    \Delta \F_2(s)&=3a_1+3b_1\,s-3d_1s^2+3e_1\,s^2\,(s-9r)\,,
\end{split}
\eeq
while for the $C$-odd contributions we find
\beq\label{eq:AmbiguityGH}
\begin{split}
    \Delta \G_1(s)&=a_0+b_0\,s+c_0\,s^{2}\,(3r-s)\,,  
    \\[.1cm]
    \Delta \H_1(s)&=a_2+b_2\,s+c_2\,s^2\,,
    \\[.1cm]
    \Delta \H_2(s)&=d_2-3a_2\,s+3b_2\,s\,(s-3r)+9c_2\,r\,s\,(s-2r)-c_2\,s^3\,.
\end{split}
\eeq
The invariance groups of the amplitudes $\M_1^{C}$, $\M_0^{\not C}$, and $\M_2^{\not C}$, given by polynomial ambiguities, are different and independent of each other (in contrast to the erroneous assumption made in Ref.~\cite{Gardner:2019nid}).

Finally, we would like to address the issue of corrections to the reconstruction theorems for $\M_{1}^{C}$, $\M_{0}^{\slashed{C}}$, and $\M_{2}^{\slashed{C}}$ stated in Eq.~\eqref{eq:reconstruction theorems}. The next discontinuities, beyond those in $S$- and $P$-waves, would come from $D$- (for even isospin) and $F$-waves (for odd isospin). Since the symmetry structure of the isoscalar amplitude does not allow for even partial waves, it is obvious that its reconstruction theorem actually holds up to corrections due to $F$- and higher (odd) partial waves. Moreover, possible $D$-wave contributions to the discontinuity of the isotensor amplitude are only allowed to have $I=2$, which is a nonresonant and extremely small partial wave at low energies.  As the validity of the reconstruction theorem for the $C$-conserving amplitude $\M_{1}^{C}$, which neglects discontinuities due to $I=0$ $D$-wave pion--pion rescattering, is well-established and tested against very accurate data, we conclude that corrections to the decomposition of the $C$-violating amplitudes $\M_{0}^{\slashed{C}}$ and $\M_{2}^{\slashed{C}}$ are necessarily even smaller and therefore entirely negligible.  

\subsection{Elastic unitarity}
\label{sec:elastic unitarity eta}
Within the scope of this work we will exclusively study the dominant elastic rescattering effects, i.e., we restrict the evaluation of the single-variable functions to $\pi\pi$ intermediate states only. In order to obtain an amplitude with manifest unitarity, each single-variable function has to obey the discontinuity relation\footnote{Note that for elastic two-body scattering the discontinuity can be replaced by the imaginary part of $\A_I(s)$. In contrast, for a three-body decay, which is obtained by analytic continuation, the right-hand side of the unitarity equation above is not purely imaginary as $\hat\A_I(s)$ becomes complex and Schwarz' reflection principle is not applicable anymore. Therefore we will solely refer to the discontinuity of $\A_I(s)$.}  
\beq\label{eq:DiscEta3Pi}
    \disc \A_I(s) = 2i\,\theta(s-4M_\pi^2)\,\big[\A_I(s)+\hat \A_I(s)\big]\,\sin\delta_I(s)\,e^{-i\delta_I(s)}\,. 
\eeq
Here we introduced the so called \textit{inhomogeneities} $\hat \A_I(s)$ that do not have a discontinuity along the right-hand cut and can be evaluated by a projection onto the respective partial wave\footnote{We define the partial-wave projection for a scalar $2\to2$ scattering amplitude $\T_I(s,z_s)$ for fixed two-particle isospin $I$ by 
\begin{equation*}
   a_I^\ell(s)\equiv a_I(s)=\frac{2\ell+1}{2\kappa^\ell(s)}\int_{-1}^1\diff z_s\, P_\ell(z_s)\,\T_I(s,z_s)\,,
\end{equation*}
where $P_\ell$ denote the Legendre polynomials. For details on how these partial-wave amplitudes relate to $\M_0^{\not C}$, $\M_1^C$, and $\M_2^{\not C}$, we refer to Ref.~\cite{Isken:2021gez}.}
\beq
    a_I(s)=\A_I(s) +\hat \A_I(s)\,.
\eeq
Note that the full information about the discontinuity of the partial wave along the right-hand cut is contained in the respective $\A_I(s)$. Let us now, for the sake of simplicity, define the angular average
\beq\label{eq:AngAver}
    \langle z_{s}^n\,\A_{I}\rangle\equiv\frac{1}{2}\int_{-1}^{1}\diff z_{s}\,z_{s}^n\,\A_{I}\big(t(s,z_{s})\big)\,.
\eeq
This allows to write the inhomogeneities for the Standard-Model amplitude in the shortened form
\beq
\begin{aligned}
    \hat \F_0(s)& =\frac{2}{9}\big[ 3\langle \F_0 \rangle +9(s-r)\,\langle \F_1 \rangle +3\kappa\,\langle z_{s}\,\F_1 \rangle +10\langle \F_2 \rangle \big]\,,
    \\[.1cm]
	\hat \F_1(s)&= \frac{1}{2\kappa}\big[ 6\langle z_{s}\,\F_0 \rangle +9(s-r)\,\langle z_{s}\,\F_1 \rangle +3\kappa\,\langle z_{s}^2\,\F_1 \rangle -10\langle z_{s}\,\F_2 \rangle \big]\,,
	\\[.1cm]
	\hat \F_2(s)&= \frac{1}{6}\big[ 6\langle \F_0 \rangle -9(s-r)\,\langle \F_1 \rangle -3\kappa\, \langle z_{s}\,\F_1 \rangle +2\langle \F_2 \rangle \big]\,,
\end{aligned}
\eeq
and the ones for the $C$-violating contributions as
\beq
\begin{aligned}
	\hat\G_1(s)&=-\frac{3}{\kappa} \big[ 3(s-r)\,\langle z_{s}\,\G_1 \rangle +\kappa\,\langle z_{s}^2\,\G_1 \rangle \big]\,,
	\\[.1cm]
	\hat\H_1(s)&=\frac{3}{2\kappa} \big[ 3(s-r)\,\langle z_{s}\,\H_1 \rangle +\kappa\,\langle z_{s}^2\,\H_1 \rangle +2\langle z_{s}\,\H_2 \rangle \big]\,,
	\\[.1cm]
	\hat\H_2(s)&=\frac{1}{2} \big[ 9(s-r)\,\langle \H_1 \rangle +3\kappa\, \langle z_{s}\,\H_1 \rangle -2\langle \H_2 \rangle \big]\,.
\end{aligned}
\eeq
Note that the argument of $\A_{I}$ in Eq.~\eqref{eq:AngAver} is Mandelstam $t$, meaning that the inhomogeneity in the $s$-channel single-variable function is determined by contributions of the crossed channels. In other words, $\hat\A(s)$ contains left-hand-cut contributions to the respective partial wave.

In order to obtain a unique solution for the discontinuity relation in Eq.~\eqref{eq:DiscEta3Pi} it is appealing to first consider the homogeneous case by setting $\hat\A_I(s)~=~0$.\footnote{This scenario is consistent with Watson's final-state theorem, which states that the phase of $\A_I$ coincides with the phase shift of elastic $\pi\pi$ rescattering.}  The homogeneous solution corresponds to the one of a pion form factor (of the appropriate quantum  numbers) and is given in terms of the Omnès function~\cite{Omnes:1958hv}  
\beq\label{eq:Omnes}
	\Omega(s)=\exp\bigg( \frac{s}{\pi} \int_{4M_\pi^2}^\infty \frac{\diff x}{x}\,\frac{\delta(x)}{(x-s)}\bigg)\,.
\eeq
Using the latter, the general solution becomes 
\beq\label{eq:DispersionIntegral}
    \A_I(s)= \Omega_I(s)\,\bigg(P_{n-1}(s)+\frac{s^n}{\pi}\int_{4M_\pi^2}^\infty \frac{\diff x}{x^{n}}\, \frac{\sin\delta_I(x)\,\hat{\A}_I(x)}{|\Omega_I(x)|\,(x-s)}\bigg)\,,
\eeq
where $P_{n-1}(s)$ is a polynomial in $s$ of order $n-1$. Its coefficients are known as subtraction constants, which are the only free parameters of our amplitude.

Throughout this paper, we will assume all subtraction constants within the same decay amplitude representation to be \textit{relatively real}.  This is not rigorously true to arbitrary precision, as the SVAs do not fulfill the Schwarz reflection principle, and their discontinuities are complex.  However, the potential imaginary parts of the subtraction constants scale with the available three-body phase space, and therefore are tiny for decays such as $\eta\to3\pi$ or $\eta'\to\eta\pi\pi$.  This has been tested explicitly for $\eta\to3\pi$~\cite{Colangelo:2018jxw}, making use of the two-loop representation in chiral perturbation theory~\cite{Bijnens:2007pr}, with the result that imaginary parts in the dispersive subtractions are entirely negligible.  This is, however, not true any more for Khuri--Treiman representations of three-body decays with larger energy releases, see $\phi\to3\pi$~\cite{Niecknig:2012sj} or certain $D$-meson decays~\cite{Niecknig:2015ija,Niecknig:2017ylb}.

Besides these degrees of freedoms, which have to be fixed by data regression or matching to effective theories, the only input for the dispersive $\eta\to3\pi$ amplitude are the $\pi\pi$ scattering phase shifts $\delta_I(s)$. 
\subsection{Subtraction scheme}
\label{sec:SubSchemeEta}
Choosing the number of subtraction constants $n$ is a rather sensitive issue. Having a purely mathematical look at the dispersion integral in Eq.~\eqref{eq:DispersionIntegral}, the minimal number is the one at which convergence is ensured. Any additional subtraction just leads to a rearrangement of the equation. The thereby introduced subtraction constants have to fulfill the corresponding sum rule, such that the minimally and higher subtracted integrals are analytically the same. Any deviation from the respective sum rule violates the initially assumed high-energy behavior and is inconsistent as a matter of principle. But allowing the additional subtraction constants to vary from the sum rule suppresses the hardly-constrained high-energy behavior of the dispersion integral and introduces additional degrees of freedom in a fit to experimental data. 

Past studies in the same Khuri--Treiman framework as presented here~\cite{Colangelo:2016jmc,Colangelo:2018jxw} put their main focus on maximal precision of the low-energy representation of the $\eta\to3\pi$ Standard-Model decay amplitude, and therefore incorporated a rather generous number of subtraction constants.  Our aim here is slightly different: we will demonstrate that with rigorous assumptions on the high-energy behavior, and accordingly a minimal number of free parameters, we are still able to describe the Dalitz plot data sufficiently well.  Subsequently, we impose the \textit{same} high-energy asymptotics on the two $C$-violating amplitudes, and show that as a result, they can be written in terms of one single subtraction constant each.  In this manner, we can prove that the mere assumption to describe the BSM amplitudes in terms of a multiplicative normalization only~\cite{Gardner:2019nid} can be justified more rigorously in terms of their analytic properties.

\begin{figure}[t!]
	\centering
	\hspace{-.75cm}\scalebox{.65}{\input{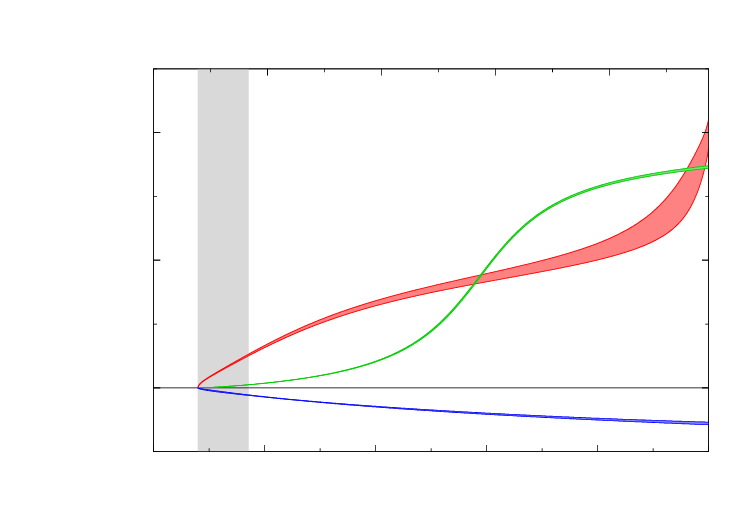}}\hspace{-.75cm}\hfill
	\scalebox{.65}{\input{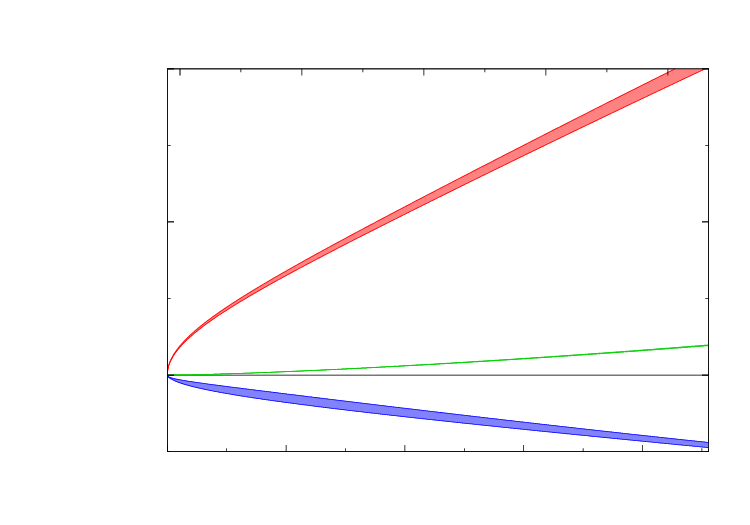}}
	\caption{The $S$- and $P$-wave $\pi\pi$ scattering phase shifts $\delta{}_{0}$ (red), $\delta_{1}$ (green), and $\delta_{2}$ (blue) covering low-energy uncertainty bands as determined by Roy equation analysis~\cite{Colangelo:2001df,Caprini:2011ky}. Left panel: behavior of the phase shifts in the low-energy region below the $K\bar{K}$-threshold at about $50\mpi^{2}$. The phase space for $\eta\to3\pi$ is indicated by the gray region. Right panel: magnification of the physical decay region.}
	\label{fig:phase shifts}
\end{figure}

In order to investigate the convergence of the dispersion integral, some assumptions have to be made for the asymptotics of $\delta_I(s)$ and $\A_I(s)$. We rely on a Roy equation analysis~\cite{Colangelo:2001df,Caprini:2011ky} to fix our phase shifts very precisely in the low-energy range, i.e., below about $1\text{GeV}^2$, as shown in Fig.~\ref{fig:phase shifts}. 
Unfortunately the high-energy behavior is not severely restricted by these equations. Therefore we suppose that in the limit $s\to\infty$ the phase shifts approximate constants
\beq
    \delta_0(s)\to \pi\,, \qquad \delta_1(s)\to \pi\,, \qquad \delta_2(s)\to 0\,,
\eeq
and analytically continue them accordingly. These limits directly fix the asymptotics of the Omnès functions, which behave for $s\to\infty$ like $s^{-k}$ if $\delta_I(s)\to k\pi$. Further, in order to use the minimal number of subtraction constants, we assume that our amplitudes scale in the limit of large momenta as 
\beq
    \A_0(s)=\mathcal{O}(s^0)\,,\qquad 
    \A_1(s)=\mathcal{O}(s^{-1})\,,\qquad
    \A_2(s)=\mathcal{O}(s^0)\,,
\eeq
and are thus even more restrictive than suggested by the Froissart--Martin bound~\cite{Froissart:1961ux}. Finally, with our minimal subtraction scheme we can obtain for the $C$-conserving Standard-Model amplitude 
\beq\label{eq:SubSchemeEta3PiC}
\begin{aligned}
            \F_0(s)&= \Omega_0(s)\,\bigg(\alpha+\beta\,s+\frac{s^2}{\pi}\int_{4M_\pi^2}^\infty\frac{\diff x}{x^{2}}\, \frac{\sin\delta_0(x)\,\hat{\F}_0(x)}{|\Omega_0(x)|\, (x-s)}\bigg)\,,\\[.1cm]
			\F_1(s)&= \Omega_1(s)\,\bigg(\gamma+\frac{s}{\pi}\int_{4M_\pi^2}^\infty	\frac{\diff x}{x}\, \frac{\sin\delta_1(x)\,\hat{\F}_1(x)}{|\Omega_1(x)|\, (x-s)}\bigg)\,,\\[.1cm]
			\F_2(s)&= \Omega_2(s)\,\bigg(\frac{s}{\pi}\int_{4M_\pi^2}^\infty \frac{\diff x}{x}\, \frac{\sin\delta_2(x)\,\hat{\F}_2(x)}{|\Omega_2(x)|\, (x-s)}\bigg)\,.
\end{aligned}
\eeq
This representation hence depends on three free parameters (all of which are chosen to be real), where Refs.~\cite{Colangelo:2016jmc,Colangelo:2018jxw} employed six.
Similarly rigorous schemes with few parameters have previously been suggested in Refs.~\cite{Guo:2015zqa,Guo:2016wsi,Albaladejo:2017hhj}.
Analogously the $C$-violating contributions become
\beq\label{eq:SubSchemeEta3PiCvio}
\begin{aligned}
			\G_1(s)&= \Omega_1(s)\,\bigg(\varepsilon+\frac{s}{\pi}\int_{4M_\pi^2}^\infty\frac{\diff x}{x}\, \frac{\sin\delta_1(x)\,\hat{\G}_1(x)}{|\Omega_1(x)|\, (x-s)}\bigg)\,,\\[.1cm]
			\H_1(s)&= \Omega_1(s)\,\bigg(\vartheta+\frac{s}{\pi}\int_{4M_\pi^2}^\infty\frac{\diff x}{x}\, \frac{\sin\delta_1(x)\,\hat{\H}_1(x)}{|\Omega_1(x)|\, (x-s)}\bigg)\,,\\[.1cm]
			\H_2(s)&= \Omega_2(s)\,\bigg(\frac{s}{\pi}\int_{4M_\pi^2}^\infty	\frac{\diff x}{x}\, \frac{\sin\delta_2(x)\hat{\H}_2(x)}{|\Omega_2(x)|\, (x-s)}\bigg)\,.
\end{aligned}
\eeq
Note that these representations are not unique. We exploited the ambiguity of the dispersive representation, as given in Eqs.~\eqref{eq:AmbiguityF} and~\eqref{eq:AmbiguityGH}, to express the single-variable functions in terms of independent subtraction constants only. Conventionally, we shifted the polynomials $\Delta\A_I$ such that the $I=2$ amplitudes do not contain subtraction constants. 
Further, we would like to remark that the normalization of each amplitude of total isospin in Eq.~\eqref{eq:FullAmpEta3Pi} has a phase that is fixed unambiguously by $T$ violation and hermiticity.
Hence, the subtraction constants $\varepsilon$ and $\vartheta$, which absorb these normalizations, are complex quantities with a fixed phase, resulting in a total of five degrees of freedom for $\M$. We furthermore note that, were it not for strong rescattering phases, interference between the (then purely real) Standard-Model and the (purely imaginary) $C$-violating amplitudes would vanish altogether, and no Dalitz-plot asymmetries would be generated at all.  This further increases the importance to implement these rescattering effects via the corresponding phase shifts exactly, using dispersion theory.\footnote{At this point we note the erroneous assumption made in Ref.~\cite{Gardner:2019nid} and the first version of this work, who both allowed for arbitrary phases between the normalizations of the SM decay amplitudes and the $C$- and $CP$-violating ones.}
The fact that these two subtraction constants indeed merely serve as overall normalizations of the BSM contributions becomes evident when realizing that  the dispersive representation is linear in the subtraction constants. This very powerful property allows us to write 
\beq
\M(s,t,u)=\sum_{\nu}\nu\,\M^{\nu}(s,t,u)\,,\qquad \M^{\nu}(s,t,u)=\M(s,t,u)|_{\nu=1,\,\mu=0, \,\ldots}\,,
\eeq
where $\nu$ and $\mu$ denote generic subtraction constants and $\M\in\{\M_0^{\not C}, \M_1^{C}, \M_2^{\not C}\}$. This procedure simplifies the numerical computation tremendously, as the corresponding \textit{basis amplitudes} $\A_I^{\nu}$, which obey analogous relations as the $\M^{\nu}$ in the equation above, can be evaluated once and for all before fixing the subtraction constants.

\subsection{Taylor invariants}
\label{sec:TaylorInvariants Eta}
Any interpretation of subtraction constants, which do not have any physical meaning on their own, should be made with caution, as they depend on the chosen subtraction scheme, on the ambiguities of the dispersive representation, and on the not-well-restricted high-energy behavior of the dispersion integrals. Changes in any of the listed aspects are absorbed in the subtraction constants when fitting to data.\footnote{Even a simple estimation of the relative and overall size of the two $C$-odd amplitudes from the subtraction constants $\varepsilon$ and $\vartheta$, as in similar fashion assumed by Ref.~\cite{Gardner:2019nid}, may be misleading. Notice that an apparent difference in these coefficients can be due to the compensation of the relative, arbitrary normalization of the basis solutions $\G_I^{\varepsilon}$ and $\H_I^{\vartheta}$ when comparing to data.}

To overcome these issues we follow the idea of Refs.~\cite{Colangelo:2016jmc,Colangelo:2018jxw}, where certain linear combinations of the subtraction constants for the SM contribution were introduced, which are identified as so-called Taylor invariants. To access those, the single-variable amplitudes $\A_I\in\{\F_{I},\G_{I},\H_{I}\}$ are expanded around $s=0$, i.e.,
\beq
    \A_I(s)=A{}_{I}^{\A}+B{}_{I}^{\A}\,s+C{}_{I}^{\A}\,s^2+D{}_{I}^{\A}\,s^3+\ldots\,.
\eeq
Inserting the series into the reconstruction theorem for the SM amplitude, cf.\ Eq.~\eqref{eq:reconstruction theorems}, one obtains\footnote{For simplicity, here and in the following we denote the order of neglected higher-order polynomial terms by $\Order(p^{2n})$, which should not be confused with the counting scheme of the chiral expansion that may include nonanalytic dependencies on quark masses etc.}
\begin{equation}
\begin{aligned}
	\mathcal{M}_{1}^{C}(s,t,u) &= F_{0}+F_{1}\,(2s-t-u)+F_{2}\,s^{2}+F_{3}\,\big[(s-t)\,u+(s-u)\,t\big]+\mathcal{O}(p^{6})\,,
\end{aligned}
\end{equation}
with the Taylor invariants
\begin{equation}
\begin{aligned}
	F_{0} &=  A{}_{0}^{\F}+r\,B{}_{0}^{\F}+\frac{4}{3}\big(A{}_{2}^{\F}+r\,B{}_{2}^{\F}\big)\,, & F_{1} &= \frac{1}{3}B{}_{0}^{\F}+A{}_{1}^{\F}-\frac{5}{9}B{}_{2}^{\F}-3r\,C{}_{2}^{\F}\,,\\[.1cm]
	F_{2} &= C{}_{0}^{\F}+\frac{4}{3}C{}_{2}^{\F}\,, & F_{3} &= B{}_{1}^{\F}+C{}_{2}^{\F}\,.
	\label{eq:taylor-invariants-isovector-amp}
\end{aligned}
\end{equation}
These can be used as theory constraints to the SM amplitude when considering that one-loop $\chi$PT~\cite{Gasser:1984pr,Colangelo:2018jxw} predicts them to be
\begin{equation}
	F_{0} = 1.176(53)\,,\quad f_{1} = 4.52(29)\GeV^{-2}\,,\quad f_{2} = 16.4(4.9)\GeV^{-4}\,,\quad f_{3} = 6.3(2.0)\GeV^{-4}\,,
	\label{eq:ChPT-taylor-invariants}
\end{equation}
where $F_0$ was used as an overall normalization by means of $f_{i}\equiv F_{i}/F_{0}$ and will furthermore serve to normalize $\M_1^C$.\footnote{In principle one can also define Taylor invariants for the SM amplitude at the two-loop level in $\chi$PT~\cite{Bijnens:2007pr,Kampf:2019bkf}. However, as demonstrated in the analysis of Ref.~\cite{Colangelo:2018jxw}, a high-precision matching requires a more flexible dispersive amplitude (i.e., more than three subtraction constants). Aside of this small flaw, we will demonstrate that our dispersive representation of the SM amplitude describes both the experimental Dalitz-plot distribution and the one-loop chiral constraints very well.}

We now apply the same strategy to the $C$-odd contributions.
The effective BSM operators of Eq.~\eqref{eq:BSM-operators}, which arise from elementary considerations such as crossing symmetry and the correct behavior under time reversal, demand the amplitudes for the $\Delta I=0$ and $\Delta I=2$ transitions at lowest contributing order to be of the form
\beq\label{eq:BSMCouplings}
\begin{aligned}
    \M^{\not C}_0(s,t,u)&=i\,g_0\,(s-t) (u-s) (t-u)+\mathcal{O}(p^8)\,,\\[.1cm]
    \M^{\not C}_2(s,t,u)&=i\,g_2\,(t-u)+\mathcal{O}(p^4)\,,
\end{aligned}
\eeq
where the couplings have the dimensions $[g_0]=\text{GeV}^{-6}$ and $[g_2]=\text{GeV}^{-2}$, respectively. 
It has to be remarked that this simple polynomial expansion is by far less accurate than the full dispersive representation, but allows one to match the couplings in a convenient way. Reproducing the structure in Eq.~\eqref{eq:BSMCouplings} with the Taylor series from above we obtain
\beq\label{eq:g0g2couplings}
    g_0=i\,\varepsilon\,\big(C{}_{1}^{\G^\varepsilon} +3r\,D{}_{1}^{\G^\varepsilon}\big)\,,\qquad
    g_2=i\,\vartheta\,\big(3A{}_{1}^{\H^\vartheta}+3r\,B{}_{1}^{\H^\vartheta}+B{}_{2}^{\H^\vartheta}+2r\,C{}_{2}^{\H^\vartheta}\big)\,,
\eeq
which we wrote in explicit dependence on the subtraction constants using the Taylor invariants for the basis amplitudes $\A_I^{\nu}$, by means of $C{}_{1}^{\G}=\varepsilon\, C{}_{1}^{\G^\varepsilon}$ etc.
In this form, $T$ violation demands the coupling constants $g_0$ and $g_2$ to be \textit{real}-valued. To satisfy this condition, the subtraction constants must be proportional to the complex conjugate of the linear combinations of Taylor invariants, by means of 
\beq\label{eq:subcon_phase_eta}
    i\,\varepsilon\overset{!}{=}c_\varepsilon\,\big(C{}_{1}^{\G^\varepsilon} +3r\,D{}_{1}^{\G^\varepsilon}\big)^\ast\,, \quad\text{with}\quad c_\varepsilon\in\mathds{R}\,,
\eeq
as an example for $\varepsilon$. While this condition fixes the phase of $\varepsilon$, the constant $c_\varepsilon$ is left as the only degree of freedom. We proceed analogously with $\vartheta$. 
As we extract the Taylor invariants by an expansion of the single-variable amplitudes around $s=0$, we are well below the dipion threshold, such that the contributions of the dispersion integrals 
are negligible for the decay at hand. Consequently, we can drop the real part of the subtraction constants, which have no visible effects on observables.

\subsection{Fixing the subtraction constants}
\label{sec:FixingSubcons Eta}
Once the basis solutions $\A_I^\nu$ for the Khuri--Treiman coupled integral equations are evaluated numerically, one can determine the free parameters of our dispersive representation for $\eta\to3\pi$. In summary we have the subtraction constants $\alpha$, $\beta$, $\gamma$ for the SM amplitude, where one of these can be seen as an overall normalization, as well as $\varepsilon$ fixing the $C$-violating isoscalar contribution and $\vartheta$ for the isotensor one.

To determine these degrees of freedom we employ a $\chi^{2}$-regression to three different data sets:
\begin{itemize}
    \item the Dalitz-plot distribution of $\eta\to\pi^{+}\pi^{-}\pi^{0}$ from the KLOE-2 collaboration~\cite{Anastasi:2016cdz},
    \item the Dalitz-plot distribution of $\eta\to 3\pi^{0}$ from the A2 collaboration~\cite{Prakhov:2018tou}, and
    \item the Taylor invariants of $\M_{1}^{C}$ from one-loop $\chi$PT.
\end{itemize}
Note that the latter two only address the three free parameters of the SM amplitude. However, these two data sets help to fix the relative phases between the contributions of different total isospins, and furthermore any shift in $\M^C_1$ may affect the BSM contributions when additionally comparing to the data of Ref.~\cite{Anastasi:2016cdz}.

Let us first turn our attention to the experimental data sets from the KLOE-2 and A2 collaborations.
The KLOE-2 collaboration provides the world's highest statistics for the measurement of the Dalitz-plot distribution in $\eta\to\pi^+\pi^-\pi^0$. The data distributes about $4.6\cdot 10^6$ events over 371 bins, where all bins overlapping with the physical boundaries were discarded. On the other hand, the A2 collaboration provides altogether 441 bins for a single Dalitz-plot sextant, exploiting the symmetry of $\eta\to3\pi^{0}$, and accepts bins overlapping with the phase space boundary. 
It supersedes many earlier experiments on $\eta\to3\pi^{0}$, which mostly concentrated on the leading nontrivial Dalitz-plot slope parameter~\cite{Bashkanov:2007aa,WASA-at-COSY:2008rsh,CrystalBallatMAMI:2008pqf,CrystalBallatMAMI:2008cye,KLOE:2010ytm,BESIII:2015fid}.
Let us refer to the experimental Dalitz-plot distributions by $\mathcal{D}^{\text{exp}}_{c,n}$, where the index denotes the charged or neutral channel, respectively. The binning is given, as commonly done, in terms of the dimensionless and symmetrized coordinates $x_{c,n}^{i},y_{c,n}^{i}$, where the additional index denotes the $i$-th bin at its center. These explicitly read 
\beq\label{eq:XYdef}
x_{c,n} = 
\frac{\sqrt{3}}{2 M_\eta Q_{c,n}} (u_{c,n}-t_{c,n})\,,\qquad
y_{c,n} = 
\frac{3}{2 M_\eta Q_{c,n}}\,\big[(M_\eta-M_{\pi^0})^2-s_{c,n}\big]-1\,, 
\eeq
where $Q_c = M_\eta - 2 M_{\pi^+} - M_{\pi^0}$ and $Q_n = M_\eta - 3 M_{\pi^0}$. The indices labeling the Mandelstam variables correspond to the respective kinematic map given in Ref.~\cite{Colangelo:2018jxw}.

We compare the experimental measurements to the dispersive Dalitz-plot distributions by integrating our amplitudes $\M_{c,n}$ over the respective bin
\begin{equation}
	\mathcal{D}_{c,n}^{\text{DR}}(x_{c,n}^{i},y_{c,n}^{i}) = \int_{\text{bin}\,\#i}\hspace{-.5cm}\diff x_{c,n}\,\diff y_{c,n}\,|\M_{c,n}(x_{c,n},y_{c,n})|^{2}\,.
\end{equation}
The discrepancy functions $\chi^2_{c,n}$ for the charged and neutral data sets are then defined by
\begin{equation}
	\chi^{2}_{c,n} = \sum_{i}\bigg(\frac{\mathcal{D}^{\text{exp}}_{c,n}(x_{c,n}^{i},y_{c,n}^{i})-|\M_{c,n}(x_{c,n}^{i},y_{c,n}^{i})|^2}{\Delta \mathcal{D}^{\text{exp}}_{c,n}(x_{c,n}^{i},y_{c,n}^{i})}\bigg)^{2}\,.
\end{equation}
To build in theory constraints on the Taylor invariants of the SM amplitude from one-loop $\chi$PT we introduce~\cite{Colangelo:2018jxw}
\begin{equation}
	\chi^{2}_{0} = \sum_{i=1}^3\bigg(\frac{f_i^{\chi\text{PT}}-\Re f_i}{\Delta f_{i}^{\chi\text{PT}}}\bigg)^{2}\,,
\end{equation}
where the $f_i^{\chi \text{PT}}$ denote the theoretical predictions listed in Eq.~\eqref{eq:ChPT-taylor-invariants}. To define a real-valued discrepancy function we restrict our analysis to the real parts of the $f_i$ and discuss the effects of their imaginary parts in Sect.~\ref{sec:SM constraints eta}.

When carrying out the combined regression to all three data sets we minimize the combined discrepancy function
\beq
	\chi_{\text{tot}}^{2} = \chi_{0}^{2} + \chi_{c}^{2} + \chi_{n}^{2}\,
	\label{eq:total-chisq-eta-3pi-c-violation}
\eeq
and fix the normalization of $\M_c$ such that it reproduces the Taylor invariant $F_0$ from Eq.~\eqref{eq:ChPT-taylor-invariants}. 
Before fixing the subtraction constants, one has to consider higher-order isospin corrections due to the mass difference of the neutral and charged pions in the final state to obtain an accurate description of the experimental measurements. To this end, we follow the same strategy as proposed in Ref.~\cite{Colangelo:2018jxw}, which shall serve as a reference for explicit formulas. The dominant isospin-breaking contribution can be taken into account by a kinematic map, such that the boundaries of the Dalitz plot in the isospin limit are mapped to the ones for physical masses.  All remaining isospin-breaking effects are assumed to be mostly absorbed by electromagnetic correction factors $\mathcal{K}_{c,n}$ for the charged and neutral decay modes of $\eta\to3\pi$ resulting from one-loop representations in $\chi$PT~\cite{Ditsche:2008cq}. While the kinematic map will be applied to both the $C$-even and $C$-odd amplitudes, $\mathcal{K}_{c,n}$ only enter the SM amplitudes, as we are yet missing any effective theory to account for analogous corrections in the $C$-violating amplitudes.  Due to the absence of $I=0$ $S$-wave contributions, in general we expect such electromagnetic effects to be even smaller in that case.

When minimizing the $\chi^{2}$ as described above,  we distinguish between the four scenarios
\begin{itemize}
    \item SM$_c$: \ \ \ \! exclusively minimize $\chi^2_c$ with $\M^{\not C}_{0,2}=0\,$,
    \item BSM$_c$: \ \! exclusively minimize $\chi^2_c$ with the full amplitude $\M_c\,$,
    \item SM$_\text{tot}$: \ \ minimize $\chi^2_\text{tot}$ with $\M^{\not C}_{0,2}=0\,$,
    \item BSM$_\text{tot}$: minimize $\chi^2_\text{tot}$ with the full amplitude $\M_c\,$.
\end{itemize}
A summary of the individual $\chi^{2}$ contributions to the four scenarios is given in Table~\ref{tab:chisq-results-eta-3pi-c-violation}. 
\begin{table}
\centering
\renewcommand{\arraystretch}{1.5}
\begin{tabular}{lcccccc}
\toprule
& $\chi_{0}^{2}$ & $\chi_{c}^{2}$ & $\chi_{n}^{2}$ & dof & ~$\chi_{\text{tot}}^{2}/\text{dof}$~ & ~$p$-value\\
\midrule
SM$_c$~~ & ~(1.222)~ & ~387.8~ & ~(509.5)~ & ~368~ & 1.054 & ~$22.9\%$ \\
BSM$_c$~~ & ~(1.222)~ & ~383.5~ & ~(509.5)~ & ~366~ & 1.048 & ~$25.1\%$ \\
SM$_\text{tot}$~~ & ~1.247~ & ~387.9~ & ~509.3~ & ~811~ & 1.108 & ~$\phantom{2}1.7\%$ \\
BSM$_\text{tot}$~~ & ~1.247~ & ~383.6~ & ~509.3~ & ~809~ & 1.105 & ~$\phantom{2}2.0\%$ \\
\bottomrule
\end{tabular}
\renewcommand{\arraystretch}{1.0}
\caption{Summary of the four considered fit scenarios: SM$_c$ (exclusive, $C$-conserving), BSM$_c$ (exclusive, $C$-violating), SM$_\text{tot}$ (combined, $C$-conserving), BSM$_\text{tot}$ (combined, $C$-violating).  For fits obtained by dropping the contributions of $\chi^{2}_{0}$ and $\chi_{n}^{2}$ to the total discrepancy function $\chi_{\text{tot}}^{2}$, their values are put in brackets. All values refer to fit results of our central solution.}
\label{tab:chisq-results-eta-3pi-c-violation}
\end{table}

We find for all fit scenarios considered a good agreement of our dispersive amplitude with data. Overall the individual parts of the discrepancy function $\chi_{0}^{2}$, $\chi_{c}^{2}$, and $\chi_{n}^{2}$ in the four different scenarios are almost identical. In fact, the dispersive representation is already perfectly fixed by the KLOE-2 data on $\eta\to\pi^{+}\pi^{-}\pi^{0}$ alone, with the $\eta\to3\pi^{0}$ Dalitz-plot distribution and the Taylor invariants for $\M_{1}^{C}$ being a prediction in excellent agreement with data. Accordingly, the differences between the results for the exclusive and combined fits are marginal, i.e., comparing SM$_c$ vs.\ SM$_\text{tot}$ and BSM$_c$ vs.\ BSM$_\text{tot}$. Taking the $C$-violating contributions into account and comparing SM$_c$ vs.\ BSM$_c$ or SM$_\text{tot}$ vs.\ BSM$_\text{tot}$ respectively, we find a minor improvement of $\chi_{c}^{2}$ by about $1.1\%$, whereas $\chi_{0}^{2}$ and $\chi_{n}^{2}$ do not change at all at the given level of accuracy. Furthermore, comparing the resulting discrepancy functions of the KLOE-2 and A2 data sets, we notice a slightly worse description of the Dalitz-plot distribution for the neutral $\eta\to3\pi^{0}$ mode. This small tension of the dispersive representation for $\M_{n}$ and the experimental measurement from A2 has also been observed in Ref.~\cite{Colangelo:2018jxw}. Nevertheless, the experimental data of both the charged and neutral mode together are well described. Consequently, adding the contributions of $\M_{0}^{\slashed{C}}$ and $\M_{2}^{\slashed{C}}$ to our dispersive representation for $\M_{c}$ has no visible effect on the determination of $\M_{1}^{C}$. 

Let us now elaborate on the error analysis. For the latter we consider the experimental uncertainties from the KLOE-2 and A2 Dalitz-plot distributions,\footnote{The A2 collaboration provided us with three independent sets of their data, allowing us to assess the statistical and systematical uncertainties of their analysis. In case of the KLOE-2 data set we will consider only the statistical errors.} the uncertainty originating from $\chi$PT constraints including the Taylor invariants for $\M_{1}^{C}$ \eqref{eq:ChPT-taylor-invariants} and the electromagnetic correction factors $\mathcal{K}_{c,n}$ from Ref.~\cite{Colangelo:2018jxw}, and the uncertainty resulting from the variation of the phase shift input in the low- and high-energy region, cf.\ Fig.~\ref{fig:phase shifts}. We will treat all these sources of error as symmetric and Gaussian distributed. Accordingly, the combined total uncertainties are found by adding the individual contributions in quadrature and the presented correlation matrices are calculated from the respective total covariance matrices of the investigated quantities.

\begin{table}[t!]
	\centering
	\renewcommand{\arraystretch}{1.5}
	\setlength{\tabcolsep}{5pt}
	\setlength\extrarowheight{2pt}
	\begin{tabular}{lccccc}
		\toprule
		& $\alpha$ & $\beta\cdot M_\pi^{2}$ & $\gamma\cdot M_\pi^{2}$ & $\Im\varepsilon\cdot M_\pi^{2}$ & $\Im\vartheta\cdot 10^{3} M_\pi^{2}$ \\
 		\midrule
	SM$_\text{tot}$& $0.92(4)$ & $-0.026(3)$ & $0.096(4)$   & -- & --  \\
	BSM$_\text{tot}$& $0.92(4)$ & $-0.026(3)$ & $0.096(4)$ & $0.014(22)$ & $0.068(34)$ \\
	\bottomrule
	\end{tabular}
\renewcommand{\arraystretch}{1.0}
	\caption{Results for the subtraction constants of the SM amplitude in the first row and the full BSM representation in the second row for the fit scenarios SM$_\text{tot}$ and BSM$_\text{tot}$.}
	\label{tab:SubConsEta3Pi} 
\end{table}
\begin{figure}[t!]
 	\centering
 	\scalebox{.8}{\input{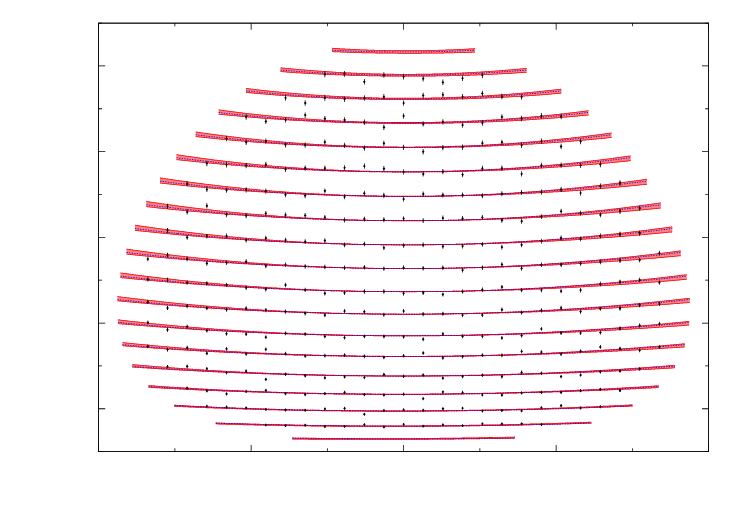}}
	\caption{Comparison of the dispersive Dalitz-plot distribution for $\eta\to\pi^{+}\pi^{-}\pi^{0}$ to experimental data. The distributions are normalized to one at the Dalitz-plot center $x_{c}=y_{c}=0$. From top to bottom we depict slices through the Dalitz plot given for $y_{c}^{\text{min}}=-0.95$ to $y_{c}^{\text{max}}=0.85$ at distances of $\Delta y_{c} = 0.1$. We show the modulus square of the full amplitude $|\M_{c}|^{2}$ with its uncertainty band covering the statistical and systematical errors added in quadrature (red) as well as the central solution for the $C$-conserving part $|\M_{1}^{C}|^{2}$ (blue). The 371 data points with error bars (black) were provided by the KLOE-2 collaboration~\cite{Anastasi:2016cdz}.}
	\label{fig:FitKLOE}
\end{figure}
For the sake of completeness we quote the subtraction constants determined with the combined regressions SM$_\text{tot}$ and BSM$_\text{tot}$ in Table~\ref{tab:SubConsEta3Pi}, which underline the findings pointed out in the previous paragraphs, and illustrate the corresponding comparison to the KLOE-2 Dalitz plot in Fig.~\ref{fig:FitKLOE}. Due to the reasons stated in Sect.~\ref{sec:TaylorInvariants Eta} we will refrain from any further discussion of the subtraction constants and instead have a look at actual observables in the following sections. Based on the observations discussed above, we will henceforth exclusively refer to the results obtained with the scenario BSM$_\text{tot}$. 

\begin{figure}[t!]
 	\centering
 	\scalebox{.8}{\input{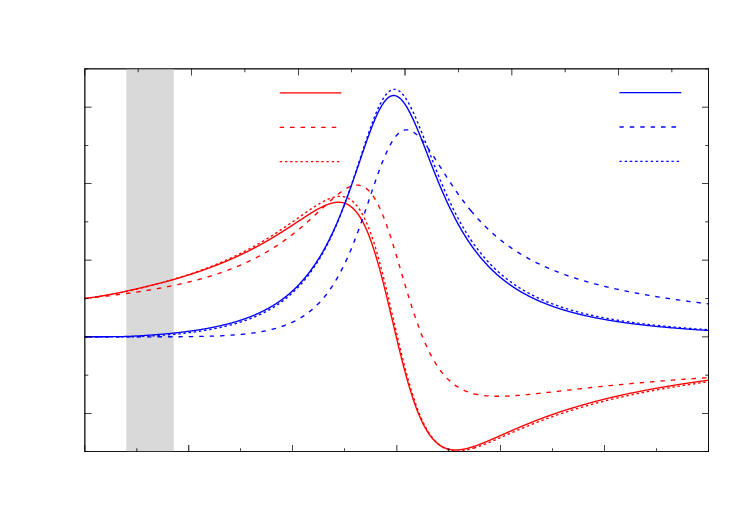}}
	\caption{$P$-wave single-variable amplitudes as defined in Eqs.~\eqref{eq:SubSchemeEta3PiC} and~\eqref{eq:SubSchemeEta3PiCvio}. Each amplitude is normalized to 1 at $s=0$ and $\F_1$ is fixed by the central values of the respective subtraction constants from Table~\ref{tab:SubConsEta3Pi}. The phase space for $\eta\to3\pi$ is indicated by the gray region.}
	\label{fig:P-wave SVAs}
\end{figure}

After fixing the $C$-conserving contribution with the subtraction constants listed in Table~\ref{tab:SubConsEta3Pi} we can compare the three contributions $\M_1^C$, $\M_0^{\not C}$, and $\M_2^{\not C}$ on the level of SVAs $\F_I$, $\G_I$, and $\H_I$. For this purpose we depict the respective normalized $P$-wave SVAs in Fig.~\ref{fig:P-wave SVAs}. With the shown extrapolation to the region of the $\rho(770)$ resonance, we observe that the isovector and isotensor contributions, i.e., $\F_1$ and $\H_1$, are in good agreement with each other over an energy range exceeding the physical decay region. On the contrary, these SVAs are significantly different from the isoscalar one, i.e., $\G_1$.\footnote{Although the effects on the physical decay range might be smaller than the extrapolations in the figure suggest.} Hence the approximation that $\F_I(s)=\G_I(s)=\H_I(s)$ as assumed in Ref.~\cite{Gardner:2019nid} may not be accurate enough to investigate possible future measurements of $C$-violating effects in $\eta\to\pi^+\pi^-\pi^0$.

\subsection{Extraction of observables}
\label{sec:Observables Eta}
In this section we present the numerical results for several observables that can be extracted from our dispersive representation of the $\eta\to 3 \pi$ amplitudes. We start our discussion by first investigating theoretical and experimental constraints imposed on the SM amplitude and focus on a comparison of our results to the established analysis of Ref.~\cite{Colangelo:2018jxw}, which shall serve as a consistency check of our dispersive representation. We show that our minimal subtraction scheme for the SM amplitude meets these requirements and can thus argue that the application of this subtraction scheme to the BSM amplitude, cf.\ Eq.~\eqref{eq:SubSchemeEta3PiCvio}, is justified.

Subsequently we have a closer look at $C$-violating observables of the $\eta\to\pi^+\pi^-\pi^0$ Dalitz-plot distribution, the occurring asymmetries, and the coupling strength of effective BSM operators with isospins $\Delta I=0$ and $\Delta I=2$.

\subsubsection{Standard Model constraints}
\label{sec:SM constraints eta}

Let us start the discussion concerning the validity of our SM amplitude $\M_1^C$ by having a look at theoretical constraints from one-loop $\chi$PT. For this purpose we extract the Taylor invariants as described in Sect.~\ref{sec:TaylorInvariants Eta}. Our value for the normalization of the Taylor invariants yields $F_{0} = 1.176(53) - 0.0094(14)\, i$. As stated previously we fixed the normalization of $\M_1^C$ so that it reproduces $\Re F_{0}$ from Eq.~\eqref{eq:ChPT-taylor-invariants}, but allowed $\Im F_0$ to vary. Since the latter is exclusively generated by contributions of the dispersion integrals to Eq.~\eqref{eq:SubSchemeEta3PiC} it is roughly two orders of magnitude smaller than $\Re F_{0}$ and will be neglected from now on. For the real parts of the reduced coefficients $f_{i}$ and their correlation we hence find
\begin{equation}
\begin{array}{rr|ccc}
	\Re f_{1}/\GeV^{-2} = & \ 4.34(15)\, & 1.00 & \phantom{+}0.24 & -0.13\\[.1cm]
	\Re f_{2}/\GeV^{-4} = & \ \!\!\!12.99(52)\, & \phantom{1.00} & \phantom{+}1.00 & \phantom{+}0.03\\[.1cm]
	\Re f_{3}/\GeV^{-4} = & \ 7.54(59)\, & \phantom{1.00} & \phantom{+1.00} & \phantom{+}1.00
\end{array}\,,
\label{eq:DR-c-onserving-taylor-invariants}
\end{equation}
which are in good agreement with the prediction of one-loop $\chi$PT as quoted in Eq.~\eqref{eq:ChPT-taylor-invariants}. In contrast to the dispersive representation of Ref.~\cite{Colangelo:2018jxw}, which uses a subtraction scheme for $\M_{1}^{C}$ involving six independent subtraction constants, our minimalist scheme \eqref{eq:SubSchemeEta3PiC} is extremely stiff. Therefore it does not allow for a large variation of the reduced Taylor invariants, cf.\ Table~\ref{tab:chisq-results-eta-3pi-c-violation}. Similar to $\Im F_{0}$ the imaginary parts of the reduced invariants $\Im f_{1} = 0.193(29)\,\text{GeV}^{-2}$, $\Im f_{2} = -0.006(85)\,\text{GeV}^{-4}$, and $\Im f_{3} = -0.128(39)\,\text{GeV}^{-4}$ are found to be small.
\begin{figure}[t!]
	\centering
	\scalebox{.8}{\input{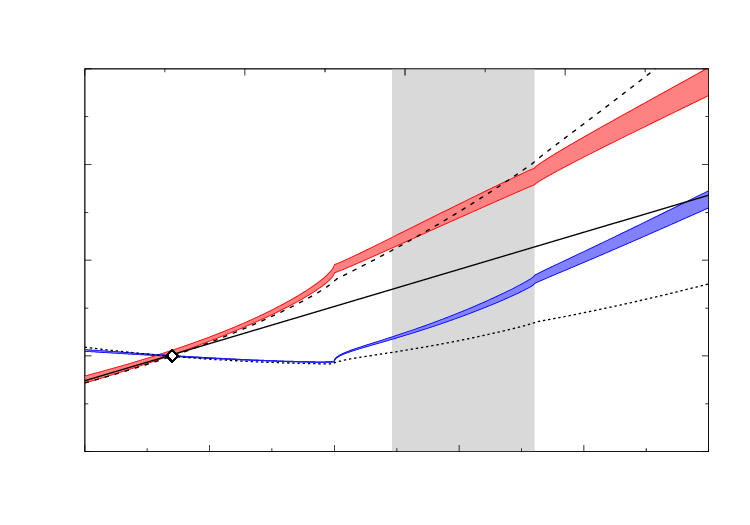}}
	\caption{Comparison of the dispersive amplitude $\mathcal{M}^{C}_{1}$ with the respective tree and one-loop level expressions obtained from $\chi$PT along the critical line $s=u$. The real and imaginary parts of the dispersive amplitude are given by the red and blue bands, which cover the range of statistical and systematical uncertainties added in quadrature. The tree level result is depicted by the solid black line, while the one-loop real part is given by the dashed and the imaginary part by the dotted black lines. The black open diamond denotes the position of the Adler zero in one-loop $\chi$PT. The physically allowed region for the $\eta\to3\pi$ decay is depicted by the gray area.}
	\label{fig:Adler-zeros-eta-3pi-c-violation}
\end{figure}
Next, we want to consider the behavior of $\M_{1}^{C}$ at its soft-pion point, i.e., in the limit where the four-momentum of one of the pions vanishes. As current algebra dictates, the amplitude $\M_{1}^{C}$ must exhibit a zero at this point. In terms of the Mandelstam variables we will find two of these so called \textit{Adler zeros} at $s_{A}=t_{A}=0$ and $s_{A}=u_{A}=0$ related by crossing symmetry. These zeros are protected by chiral $\text{SU}(2)_{L}\times\text{SU}(2)_{R}$ flavor symmetry, hence their positions are only subject to corrections of $\mathcal{O}(M_{\pi}^{2})$ if the pion mass is turned on again. At tree level the amplitude exhibits a zero crossing at $s_{A}=\frac{4}{3}M_{\pi}^{2}$~\cite{Osborn:1970nn}. A study of one-loop $\chi$PT yields a slight shift of the Adler zero to $s_{A}\approx 1.4M_{\pi}^{2}$~\cite{Gasser:1984pr}.
In Fig.~\ref{fig:Adler-zeros-eta-3pi-c-violation} the behavior of our dispersive representation for $\mathcal{M}^{C}_{1}$ along the critical line $s=u$ is compared to the tree level and one-loop predictions of $\chi$PT. We extract the zero crossing of the dispersive representation at
\begin{equation}
\begin{array}{rr|cc}
s_{A}/M_{\pi}^{2} = & \ 1.29(13)\, & 1.00 & -0.85\\[.1cm]
(s_{A}-t_{A})/M_{\pi}^{2} = & \ \!\!\!-0.057(15)\, & \phantom{1.00} & \phantom{+}1.00
\end{array}\,,
\label{eq:adler-zero-pos}
\end{equation}
which is in perfect agreement with the $\chi$PT prediction. Nevertheless, we want to mention that the Adler zeros are shifted slightly away from the critical lines $s=t$ and $s=u$. The dominating error source in Eq.~\eqref{eq:adler-zero-pos} stems from the low- and high-energy uncertainties of the phase shift input, cf.\ Fig.~\ref{fig:phase shifts}.

One last consistency check regards the observables of the neutral channel $\eta\to3\pi^0$. Due to the symmetry under exchange of any two Mandelstam variables we stick to the common phenomenological parameterization in terms of the polar coordinates $z_{n}$ and $\phi_{n}$ given by
\begin{equation}
	|\M_{n}(z_{n},\phi_{n})|^{2} \sim 1+2\alpha\,z_{n}+2\beta\,z_{n}^{3/2}\,\sin3\phi_{n}+\ldots\,.
\end{equation}
Performing a two-dimensional Taylor expansion of our amplitude results in
\begin{equation}
\begin{array}{rr|cc}
	\alpha = & \ \!\!\!-0.0293(31)\, & 1.00 & -0.77 \\[.1cm]
	\beta = & \ -0.0043(8)\, & \phantom{1.00} & \phantom{+}1.00 \\[.1cm]
\end{array}\,,
\label{eq:daln-params}
\end{equation}
where the slope $\alpha$ agrees well with the Particle Data Group (PDG) world average~\cite{Zyla:2020zbs} and the parameter $\beta$ is compatible with the findings of the A2 collaboration~\cite{Prakhov:2018tou} as well as with the dispersive analysis of Ref.~\cite{Colangelo:2018jxw}. The extraction of higher parameters is beyond the scope of this work. 

Finally, we can also calculate the ratio $\BR(\eta\to3\pi^{0})/\BR(\eta\to\pi^{+}\pi^{-}\pi^{0})$, which can be computed from partial decay widths $\Gamma_{c,n}$ defined by
\begin{equation}
	\Gamma_{c,n}(\eta\to3\pi) = \frac{Q_{c,n}^{2}}{384\sqrt{3}\pi^{3}M_{\eta}}\,\frac{\mathcal{D}_{c,n}}{S_{c,n}}\,,\qquad \mathcal{D}_{c,n} = \int\diff x_{c,n}\,\diff y_{c,n}\,|\M_{c,n}(x_{c,n},y_{c,n})|^{2}\,,
	\label{eq:}
\end{equation}
where $S_{c}=1$ and $S_{n}=6$ denoting the symmetry factors and $\mathcal{D}_{c,n}$ the integrals of the Dalitz-plot distributions over the full phase space. Since contributions antisymmetric under $t\leftrightarrow u$ cancel, $\mathcal{D}_{c}$ is determined entirely by $|\M_{1}^{C}|^{2}$ up to corrections quadratic in the BSM couplings. We extract
\begin{equation}
	\frac{\text{BR}(\eta\to3\pi^{0})}{\text{BR}(\eta\to\pi^{+}\pi^{-}\pi^{0})} = 1.423(48) 
	\label{eq:ratio-branching-eta-3pi-c-violation}
\end{equation}
in perfect agreement with the PDG world average~\cite{Zyla:2020zbs}. Note that the uncertainty quoted in Eq.~\eqref{eq:ratio-branching-eta-3pi-c-violation} is totally dominated by the errors on the electromagnetic correction factor $\mathcal{K}_{c,n}$ from Ref.~\cite{Colangelo:2018jxw}.

As our minimal subtraction scheme meets all the presented constraints imposed on the SM amplitude, we conclude that there is no objection when applying it to the BSM contributions.  

\subsubsection{Dalitz-plot distributions}
\label{sec:Dalitzplots Eta 3 pi}

\begin{figure}
    \centering
    \hspace{-.25cm}\scalebox{.63}{\input{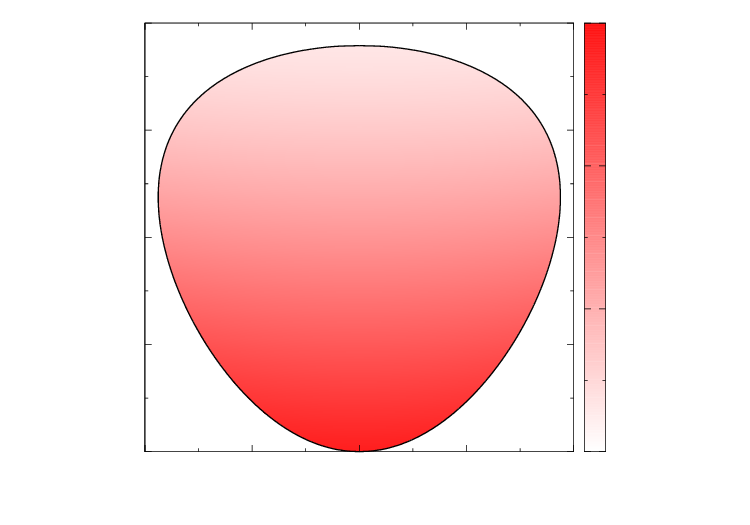}}\hspace{-.75cm}\hfill
    \scalebox{.63}{\input{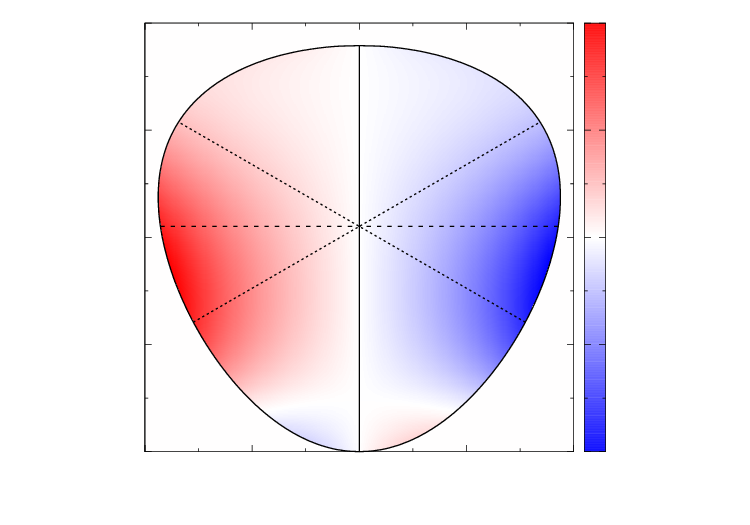}}\\
    \hspace{-.25cm}\scalebox{.63}{\input{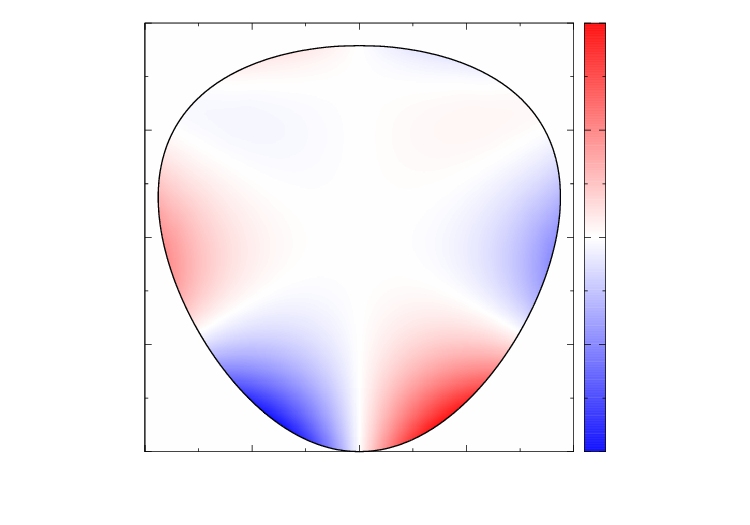}}\hspace{-.75cm}\hfill
    \scalebox{.63}{\input{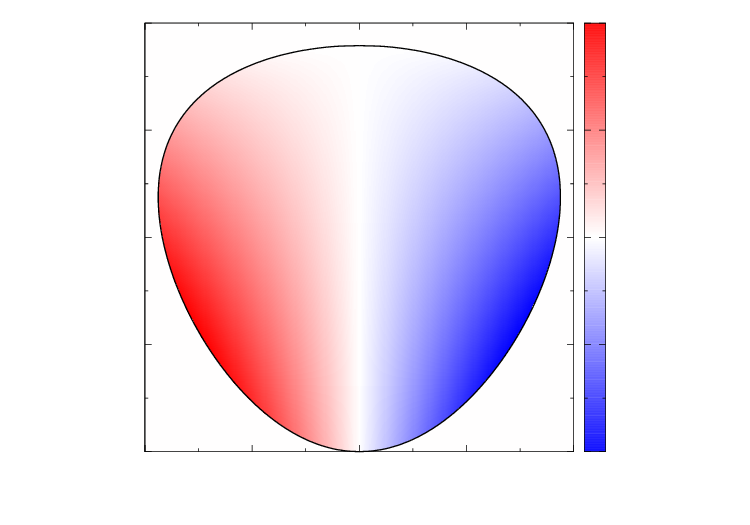}}
    \caption{Decomposition of the Dalitz-plot distribution for $\eta\to\pi^+\pi^-\pi^0$ as given in Eq.~\eqref{eq:DalitzDecompositionEta} for the central fit result. The normalization is chosen such that $|\xi\M_{1}^{C}|^2$ (top left) is one in the center. Note the individual scales of each contribution. The interferences of $\M_1^C$ with $\M_0^{\not C}$ (bottom left) and $\M_2^{\not C}$ (bottom right) give rise to mirror symmetry breaking in the Dalitz plot. The total $C$-violating contributions to the full Dalitz plot is shown in the upper right, including the symmetry axes to define asymmetry parameters. The left-right asymmetry $A_{LR}$ compares the population of the left and right halves divided by the line $t_{c}=u_{c}$ (solid vertical line), $A_{Q}$ the quadrants divided by $t_{c}=u_{c}$ and $s_{c}=r_{c}$ (solid vertical and dashed horizontal lines), and $A_{S}$ the sextants divided by $t_{c}=u_{c}$, $s_{c}=t_{c}$, and $s_{c}=u_{c}$ (solid vertical and dotted diagonal lines). The most significant impacts of the $C$-violating amplitudes are located in the vicinity of the kinematic boundary.}
    \label{fig:DalitzPlotsEta}
\end{figure}

We now turn our focus to the determination of $C$-violating observables in the $\eta\to\pi^+\pi^-\pi^0$ Dalitz-plot distribution. As already observed in Sect.~\ref{sec:FixingSubcons Eta}, patterns arising from TOPE forces have a vanishingly small influence on the goodness of the regression. Nevertheless, we show to which order of magnitude $C$-and $CP$-violating signals in $\eta\to\pi^+\pi^-\pi^0$, as predicted by our dispersive representation, can be restricted with the currently most precise measurement of the respective Dalitz plot~\cite{Anastasi:2016cdz}.
For this purpose, it may be advantageous to decompose the Dalitz-plot distribution of the total amplitude in Eq.~\eqref{eq:FullAmpEta3Pi} into its constituents by means of 
\beq\label{eq:DalitzDecompositionEta}
\begin{split}
	\big|\M_{c}\big|^{2} &\approx \big|\xi\,\M_{1}^{C}\big|^{2} + 2\Re\big[\xi\,\M_{1}^{C}\,(\M_{0}^{\slashed{C}})^{*}\big] + 2\Re\big[\xi\,\M_{1}^{C}\,(\M_{2}^{\slashed{C}})^{*}\big] \,,
\end{split}
\eeq
where we neglected all contributions that are quadratic in $C$-violating amplitudes, i.e., $|\M_{0}^{\slashed{C}}|^{2}$, $|\M_{2}^{\slashed{C}}|^{2}$, as well as $2\Re[\M_{0}^{\slashed{C}}\,(\M_{2}^{\slashed{C}})^{*}]$, and dropped the dependence on the dimensionless coordinates $x_c$ and $y_c$ for simplicity. Since we have full control on the amplitudes $\M_{1}^{C}$, $\M_{0}^{\slashed{C}}$, and $\M_{2}^{\slashed{C}}$ appearing in Eq.~\eqref{eq:DalitzDecompositionEta}, we can study their disentangled contributions to the Dalitz-plot distribution for our central fit results individually, cf.\ Fig.~\ref{fig:DalitzPlotsEta}. 
Obviously, the $C$-conserving SM part determined by $\M_{1}^{C}$ is dominating, while the two terms linear in the $C$-violating amplitudes $\M_{0}^{\slashed{C}}$ and $\M_{2}^{\slashed{C}}$ are suppressed by three orders of magnitude. We remark that all remaining terms quadratic in $C$-violating amplitudes not shown in the figure are suppressed by five to six orders of magnitude and are hence indeed totally negligible. Having a look at the two contributions linear in the $C$-violating effects, which  determine the size of the mirror symmetry breaking of the Dalitz-plot distribution under $t\leftrightarrow u$, we find both contributions to be of similar size, i.e., the interference effect of $\M_{1}^{C}$ with $\M_{0}^{\slashed{C}}$ compared to the interference $\M_{1}^{C}$ with $\M_{2}^{\slashed{C}}$. Accordingly, $\M_{0}^{\slashed{C}}$ and $\M_{2}^{\slashed{C}}$ are of the same order of magnitude. Like the SM contribution $|\M_{1}^{C}|^{2}$, all effects quadratic in $C$-violation are symmetric under $t\leftrightarrow u$ and will therefore not contribute to the mirror symmetry breaking. 

Due to the small phase space of the decay at hand the Dalitz plot is typically parameterized by a polynomial expansion around its center, by means of 
\begin{equation}
\begin{aligned}
	|\mathcal{M}_{c}(x_{c},y_{c})|^{2} &\sim 1+a\,y_{c}+b\,y_{c}^{2}+c\,x_{c}+d\,x_{c}^{2}+e\,x_{c}\,y_{c}\\[.2cm]
	&\hspace{3.5cm}+f\,y_{c}^{3}+g\,x_{c}^{2}\,y_{c}+h\,x_{c}\,y_{c}^{2}+l\,x_{c}^{3}+\ldots\,,
	\label{eq:PolynomialExpansion}
\end{aligned}
\end{equation}
where the coefficients $a$, $b$, etc., are called Dalitz-plot parameters. By now, the first seven coefficients of this phenomenological parameterization have been studied by the KLOE-2 collaboration~\cite{Anastasi:2016cdz}. Note that non-vanishing values of the coefficients $c$, $e$, $h$, and $l$ odd in $x_{c}$ would directly implicate $C$-violation in $\eta\to\pi^{+}\pi^{-}\pi^{0}$ decays. We first access the Dalitz-plot parameters for the $C$-conserving contribution, generated exclusively by $\M_{1}^{C}$, by employing a two-dimensional Taylor expansion of our amplitude $\M_1^C$, resulting in
\begin{equation}
	\begin{array}{rr|ccccc}
		a = &\ \!\!\!-1.0819(14)\, & 1.00 & -0.06 & \phantom{+}0.39 & -0.47 & -0.37\\[.1cm]
		b = &\ 0.1487(34)\, & \phantom{1.00} & \phantom{+}1.00 & \phantom{+}0.57 & -0.66 & -0.60\\[.1cm]
		d = &\ 0.088(13)\, & \phantom{1.00} & \phantom{+1.00} & \phantom{+}1.00 & -0.92 & -0.99\\[.1cm]
		f = &\ 0.1131(47)\, & \phantom{1.00} & \phantom{+1.00} &\phantom{+1.00} & \phantom{+}1.00 & \phantom{+}0.90\\[.1cm]
		g = &\ -0.068(15)\, & \phantom{1.00} & \phantom{+1.00} & \phantom{+1.00} &\phantom{+1.00} & \phantom{+}1.00
\end{array}\,.
\label{eq:dalc-params-cc}
\end{equation}
The uncertainties of the parameters $b$, $d$, and $g$ are completely driven by the variation of the phase shift input, while the uncertainties of $a$ and $f$ gain sizeable contributions from all sources of error. Similarly, for the $C$-violating Dalitz-plot parameters generated by the interference effects of $\M_{1}^{C}$ with $\M_{0}^{\slashed{C}}$ and $\M_{2}^{\slashed{C}}$ we find
\begin{equation}
\begin{array}{rr|cccc}
	c = &\ \!\!\!-0.0024(12)\, & 1.00 & -1.00 & \phantom{+}0.01 & \phantom{+}0.05\\[.1cm]
	e = &\ 0.0026(13)\, & \phantom{1.00} & \phantom{+}1.00 & -0.01 & -0.05\\[.1cm]
	h = &\ 0.0034(60)\, & \phantom{1.00} & \phantom{+1.00} & \phantom{+}1.00 & -1.00\\[.1cm]
	l = &\ \!\!\!-0.0014(21)\, & \phantom{1.00} & \phantom{+1.00} &\phantom{+1.00} &  \phantom{+}1.00
\end{array}\,.
\label{eq:dalc-params-cv}
\end{equation}
The uncertainties of these four parameters are dominated by the statistical error of the KLOE-2 data, while all other sources of uncertainty do not yield any significant contribution to the error budget.\footnote{Note that the estimated correlations between the $C$-conserving and $C$-violating parameters given in Eqs.~\eqref{eq:dalc-params-cc} and \eqref{eq:dalc-params-cv} are below $1\%$.}
Accordingly, we can confirm that all $C$-violating parameters vanish within $2\sigma$ at most. Furthermore, the $C$-violating parameters turn out to be at least one order of magnitude smaller than $d$ and $g$, which are the smallest coefficients of the $C$-conserving part of the parameterization \eqref{eq:dalc-params-cc}. Separating the individual contributions to the central values of $c$, $e$, $h$, and $l$ originating from the interference effect of $\M_{1}^{C}$ with $\M_{0}^{\slashed{C}}$ we find 
\begin{equation}
	c = +0.0000\,,\qquad e = +0.0000\,,\qquad h = +0.0037\,,\qquad l = -0.0013\,,
\end{equation}
whereas the interference of $\M_{1}^{C}$ with $\M_{2}^{\slashed{C}}$ yields
\begin{equation}
	c = -0.0024\,,\qquad e = +0.0026\,,\qquad h = -0.0003\,,\qquad l = -0.0002\,.
\end{equation}
A comparison of the extracted Dalitz-plot parameters with the results from KLOE-2 as well as the two most recent dispersive analyses on $C$-conserving $\eta\to3\pi$ decays~\cite{Albaladejo:2017hhj,Colangelo:2018jxw} are summarized in Table~\ref{tab:Dalitz-params-eta-3pi-c-violation}.
\begin{table}
\centering
\renewcommand{\arraystretch}{1.5}
\resizebox{\columnwidth}{!}{
\scalebox{0.75}{
\begin{tabular}{l c c c c c c c c c}
\toprule
& $-a$ & $b$ & $-c$ & $d$ & $e$ & $f$ & $-g$ & $h$ & $l$ \\
\midrule
KLOE-2 & $1.095(3)$ & $0.145(3)$ & $0.004(3)$ & $0.081(3)\phantom{0}$ & $0.003(3)$ & $0.141(7)$ & $0.044(9)\phantom{0}$ & $0.011(9)$ & $\phantom{-}0.001(7)$ \\
DR Orsay & $1.142\phantom{(2)}$ & $0.172\phantom{(4)}$ & -- & $0.097\phantom{(3)}\phantom{0}$ & -- & $0.122\phantom{(4)}$ & $0.089\phantom{(4)}\phantom{0}$ & -- & -- \\
DR Bern & $1.081(2)$ & $0.144(4)$ & -- & $0.081(3)\phantom{0}$ & -- & $0.118(4)$ & $0.069(4)\phantom{0}$ & -- & -- \\
this work & $1.082(1)$ & $0.149(3)$ & $0.002(1)$ & $0.088(13)$ & $0.003(1)$ & $0.113(5)$ & $0.068(15)$ & $0.003(6)$ & $-0.001(2)$ \\
\bottomrule
\end{tabular}
}
}
\renewcommand{\arraystretch}{1.0}
\caption{Comparison of the Dalitz-plot parameters obtained in different analyses of the KLOE-2 data~\cite{Anastasi:2016cdz}. The values given in the first row are obtained by a direct fit of Eq.~\eqref{eq:PolynomialExpansion} to data. The dispersive analyses from the Orsay~\cite{Albaladejo:2017hhj} and Bern groups~\cite{Colangelo:2018jxw} consider the $C$-conserving amplitude $\M_{1}^{C}$ only.}
\label{tab:Dalitz-params-eta-3pi-c-violation}
\end{table}

\subsubsection{Asymmetries and BSM couplings}
\label{sec:AsymmEta3Pi}
Besides these coefficients, we can also investigate three asymmetry parameters to quantify $C$-violating effects in the $\eta\to\pi^{+}\pi^{-}\pi^{0}$ Dalitz-plot distribution: the left-right $A_{LR}$, the quadrant $A_{Q}$, and sextant $A_{S}$ asymmetry parameters~\cite{Layter:1972aq,Lee:1965zza,Nauenberg:1965}. 
These asymmetries compare the population of the Dalitz-plot distribution in the different sectors defined by the Dalitz-plot geometry, cf.\ Fig.~\ref{fig:DalitzPlotsEta}. To quantify these asymmetries we follow Ref.~\cite{Gardner:2019nid} by defining 
\begin{equation}
\begin{split}
    A_{LR}&=\frac{N_{R}-N_{L}}{N} \,, \qquad
    A_{Q}=\frac{N_{A}-N_{B}+N_{C}-N_{D}}{N} \,, \\[.3cm]
    A_{S}&=\frac{N_\text{I}-N_\text{II}+N_\text{III}-N_\text{IV}+N_\text{V}-N_\text{VI}}{N} \,,
\end{split}
\end{equation}
with $N=N_{R}+N_{L}$ and 
\beq
    N_{\mathcal{C}}=\int_{\mathcal{C}} \diff x_c\, \diff y_c\, |\M_c(x_c,y_c)|^2
\eeq
denoting the normalized number of events for the total amplitude within each region $\mathcal{C}$. In our notation, $N_{R}$ and $N_{L}$ belong to the population for positive and negative values of $x_c$, respectively. The regions $A$, $B$, $C$, $D$ and I to VI denote the quadrants and sextants, respectively, in clockwise ordering, where $A$ is the quadrant for $x_c>0$, $y_c>0$ and I the sextant completely contained in $A$; cf.\ Fig.~\ref{fig:DalitzPlotsEta}. Carrying out each integral for our dispersive representation of $\M_{c}$ we obtain 
\begin{equation}
\begin{array}{rr|ccc}
	A_{LR} =& \ \!\!\!-7.9(4.5)\, & 1.00 & -0.82 & \phantom{+}0.34\\[.1cm]
	A_{Q}  =& 1.9(2.5)\, & \phantom{1.00} & \phantom{+}1.00 & -0.82\\[.1cm]
	A_{S}  =& 2.0(3.8)\, & \phantom{1.00} & \phantom{+1.00} & \phantom{+}1.00
\end{array}\,,
\label{eq:dalc-asym}
\end{equation}
where all three asymmetry parameters are given in units of $10^{-4}$. We find $A_{LR}$, $A_{Q}$, and $A_{S}$ in good agreement with the results reported by the KLOE-2 collaboration~\cite{Anastasi:2016cdz}. Again, there is no hint for $C$-violation as all three asymmetries are compatible with zero in not more than $1.8\sigma$. Note that the error budget in Eq.~\eqref{eq:dalc-asym} is completely dominated by the statistical uncertainties of the KLOE-2 data.\footnote{In fact, KLOE-2 reports that the systematic uncertainty of $A_{LR}$ dominates the statistical one. Like the results for $A_{Q}$ and $A_{S}$, $A_{LR}$ is therefore compatible with zero in less than $1\sigma$ if systematic effects are taken into account.}

In contrast to experimental studies of $C$-violating effects in the $\eta\to\pi^{+}\pi^{-}\pi^{0}$ Dalitz-plot distribution, which are limited to the investigation of $x_{c}$-odd coefficients of the phenomenological parameterization \eqref{eq:PolynomialExpansion} or the probe of the Dalitz-plot asymmetries, our dispersion theoretical analysis provides us with the tools to disentangle the individual contributions of $\M_{0}^{\slashed{C}}$ and $\M_{2}^{\slashed{C}}$. Furthermore, we are in the position to extract coupling strengths  $g_0$ and $g_2$ of the underlying isoscalar and isotensor BSM operators as defined  Eq.~\eqref{eq:g0g2couplings}. For our dispersive representation we obtain 
\begin{equation}
\begin{array}{rl|cc}
	g_{0}/\text{GeV}^{-6} = & -2.8(4.5)\,& 1.00 & \phantom{+}0.01\\[.1cm]
	g_{2}/10^{-3}\,\text{GeV}^{-2} = & -9.3(4.6)\, & \phantom{1.00} & \phantom{+}1.00
\end{array}\,.
\label{eq:BSM-operators-coupling-strengths}
\end{equation}
Note that for the central values we find a ratio of $|g_{0}/g_{2}|\approx10^{3}\GeV^{-4}$. This can be understood as follows.  Generically, as we have remarked above, the operator $X_{0}^{\slashed{C}}$ is kinematically suppressed compared to $X_{2}^{\slashed{C}}$ by 4 orders in the chiral expansion; this means that we would expect their coefficients to behave as $|g_0/g_2| \sim (1\GeV)^{-4}$, the scale given by the chiral symmetry breaking scale $4\pi\Fpi \approx 1.16\GeV$.  As the momenta throughout the $\eta\to3\pi$ Dalitz plot are of order $M_\pi$ (note the available phase space $M_\eta-3M_\pi \approx M_\pi$), this would lead to a relative suppression of the isoscalar transition with respect to the isotensor one of roughly $(M_\pi/1\GeV)^4 \approx 4\times 10^{-4}$. 
In fact, however, the data constrains both amplitudes including their respective coupling constants about equally, cf.\ Fig.~\ref{fig:DalitzPlotsEta}, which means that the experimental sensitivities rather behave like $|g_0/g_2| \sim M_\pi^{-4} \approx 2.6\times 10^{3}\GeV^{-4}$, in good agreement with what we observe. This behavior of the amplitudes $\M_{0}^{\slashed{C}}$ and $\M_{2}^{\slashed{C}}$ has also been observed in Ref.~\cite{Gardner:2019nid}.

Furthermore we can utilize these coupling strengths to obtain a more general representation of the Dalitz-plot asymmetries. Carrying out the phase space integrals individually for contributions involving interference effects of $\M_{0}^{\slashed{C}}$ or $\M_{2}^{\slashed{C}}$ in the Dalitz-plot distribution, we find that the asymmetry parameters \eqref{eq:dalc-asym} given in units of $10^{-4}$ are related to the BSM couplings $g_0$ and $g_2$ by 
\begin{align}\label{eq:asym-dalc-eta-3pi-c-violation-decomposition}
	A_{LR}&= -0.300\,g_{0}+0.936\,g_{2}\,,\nonumber\\[.1cm]
	A_{Q} &= \phantom{-}0.443\,g_{0}-0.336\, g_{2}\,,\\[.1cm]
	A_{S} &= -0.850\, g_{0}+0.043\, g_{2}\nonumber\,.
\end{align}
In these relations $g_{0}$ and $g_{2}$ enter in units of $1\GeV^{-6}$ and $10^{-3}\GeV^{-2}$, respectively. Equation~\eqref{eq:asym-dalc-eta-3pi-c-violation-decomposition} reveals that especially the sextant asymmetry parameter $A_{S}$ is sensitive to contributions generated by $\M_{0}^{\slashed{C}}$, while effects of $\M_{2}^{\slashed{C}}$ are suppressed.\footnote{Note, however, that this would cease to be true as soon as $g_{0/2}$ turned out to be of comparable natural order as suggested by the chiral power counting, i.e., $g_0/g_2 = \Order(1\GeV^{-4})$, in which case even the sextant asymmetry would be dominated by the isotensor contribution.} Separating for contributions of $\M_{0}^{\slashed{C}}$ or $\M_{2}^{\slashed{C}}$ to the central values of the asymmetry parameters, we find
\begin{equation}
	A_{LR} = 0.8\,,\qquad A_{Q} = -1.2\,,\qquad A_{S} = 2.4\,,
\end{equation}
for interference effects of $\M_{1}^{C}$ with $\M_{0}^{\slashed{C}}$, whereas the interference of $\M_{1}^{C}$ with $\M_{2}^{\slashed{C}}$ yields
\begin{equation}
	A_{LR} = -8.7\,,\qquad A_{Q} = 3.1\,,\qquad A_{S} = -0.4\,.
\end{equation}
Once more, all asymmetry parameters are given in units of $10^{-4}$.

To conclude the discussion of $\eta\to3\pi$, we would like to comment on the future experimental focus to set more severe bounds on $C$- and $CP$-violation. 
We disrecommend using the polynomial parameterization of the Dalitz plot from Eq.~\eqref{eq:PolynomialExpansion}, which is too inaccurate for this purpose, mostly because the order of the polynomial, i.e., the number of degrees of freedom, is not known a priori and depends strongly on the precision of the measurement. 
On the other hand, the measurement of two out of the three Dalitz-plot asymmetries is in principle sufficient to fix the two degrees of freedom in our amplitude representation of $C$ and $CP$ violation. Note however that we predict strong correlations between the three asymmetries, which would become even more significant if, as naturalness suggests, the isoscalar contribution is strongly suppressed compared to the isotensor one therein.  We therefore advocate the use of the more physical decay amplitudes with proper phase behavior in future experimental analyses.

\boldmath\subsection{Generalization to $\eta'\to3\pi$}\unboldmath
\label{sec:eta'}
As $\eta$ and $\eta'$ have largely the same quantum numbers and differ mainly due to their masses, the fundamental decay mechanisms into the $3\pi$ final states are also identical.  In the Standard Model, $\eta'\to3\pi$ is also almost exclusively due to the light-quark-mass difference, and the classification in terms of isospin amplitudes works in exactly the same way as for $\eta\to3\pi$.  Consequently, the same goes for $C$-violating decay mechanisms.  A major difference concerns only the total widths of $\eta$ and $\eta'$: while the partial widths of both mesons into three pions are of comparable size, the lifetime of the $\eta'$ is shorter by about a factor of 150, and hence the branching ratios make $\eta'\to3\pi$ relatively rare decay modes.  As a result, high-precision investigations of the corresponding Dalitz plots on the same level as for $\eta\to3\pi$, with the goal to put limits on $C$-odd effects therein, will most likely remain extremely difficult in the near future.
To date, the BESIII collaboration has investigated the decay dynamics in $\eta'\to3\pi$ most precisely,  with a determination of the respective branching ratios~\cite{BESIII:2015you},
a measurement of the $\eta'\to3\pi^0$ Dalitz plot~\cite{BESIII:2015fid},
and the first amplitude analysis for both charged and neutral final states~\cite{BESIII:2016tdb}.

Here, we merely intend to estimate the relative size between isoscalar and isotensor $C$-violating transitions in $\eta'\to\pi^+\pi^-\pi^0$: due to the significantly larger available phase space, we suspect the strong kinematic or chiral suppression of the isoscalar amplitude in $\eta\to\pi^+\pi^-\pi^0$ to be lifted to a certain extent; an expectation that will be borne out below.  As a result, despite the experimental difficulty due to the smaller branching ratio, as a matter of principle $\eta'\to\pi^+\pi^-\pi^0$ will be much more sensitive to the isoscalar $C$-odd operators.
For the purpose of this qualitative investigation, it is sufficient to consider a dispersive representation of the $\eta'\to\pi^+\pi^-\pi^0$ decay amplitude 
as a rescaled version of $\eta\to\pi^+\pi^-\pi^0$, 
with the mass of the $\eta$ replaced by the one for its heavier version $\eta'$, hence increasing the available phase space. 
We omit the incorporation of any inelasticities, like via the dominant decay channel $\eta'\to\eta\pi\pi$.

For the purposes of our rather qualitative argument,
we only investigate the phase space distributions of the $C$-odd contributions, because both amplitudes $\M_0^{\not C}$ and $\M_2^{\not C}$ only depend on one complex subtraction constant each. Since both amplitudes are driven by the same type of $C$- and $CP$-violating operators as the corresponding ones in $\eta\to\pi^+\pi^-\pi^0$, we suppose for our qualitative estimation that the respective coupling constants $g_0$ and $g_2$ are equal in both decays. Under this assumption we adjust the normalization of $\M_0^{\not C}$ and $\M_2^{\not C}$ in $\eta'\to\pi^+\pi^-\pi^0$ in terms of the subtraction constants $\varepsilon$ and $\vartheta$ to $g_0$ and $g_2$ as extracted from the central results of the BSM couplings in Eq.~\eqref{eq:BSM-operators-coupling-strengths}.
Note that the contribution of the dispersion integral to the real part of the subtraction constants is in this case not negligible due to the increased phase space. Hence we fix the subtraction constants according to Eq.~\eqref{eq:subcon_phase_eta}.
\begin{figure}
    \centering
    \hspace{-.25cm}\scalebox{.63}{\input{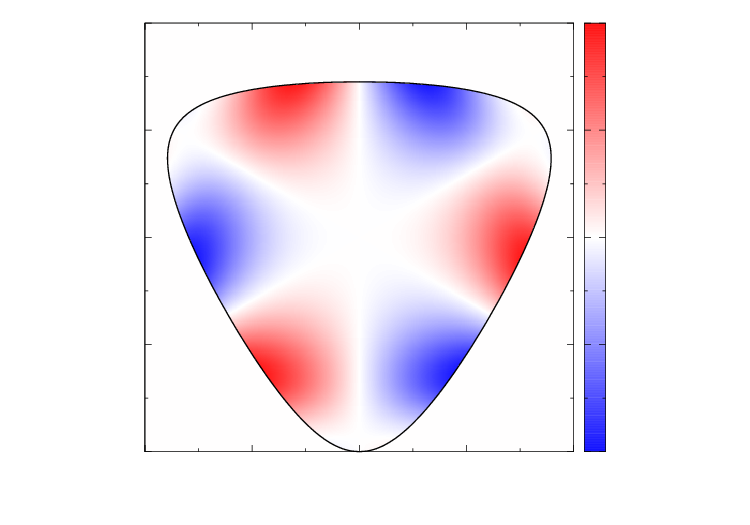}}\hspace{-.75cm}\hfill
    \scalebox{.63}{\input{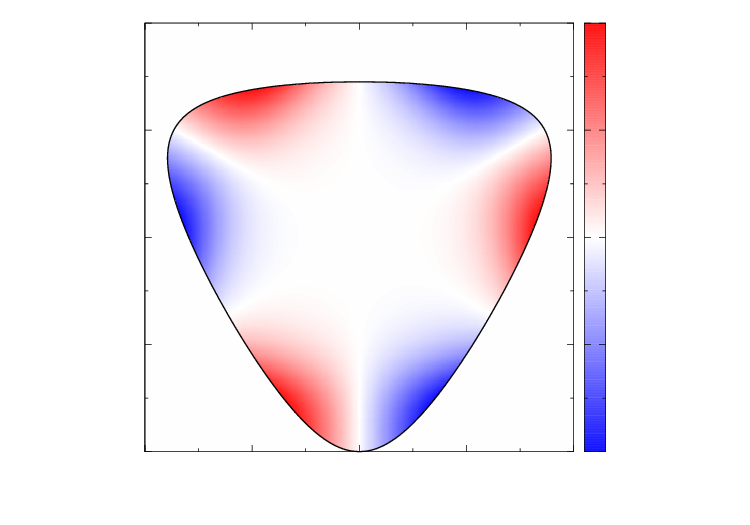}}\\
    \hspace{-.25cm}\scalebox{.63}{\input{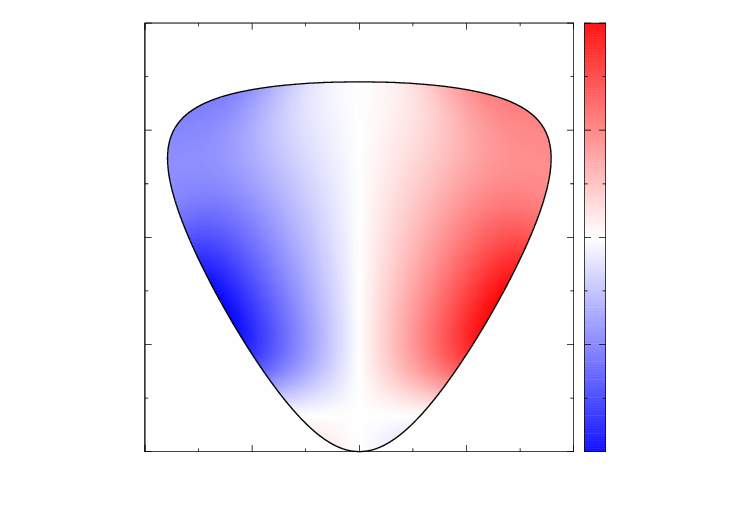}}\hspace{-.75cm}\hfill
    \scalebox{.63}{\input{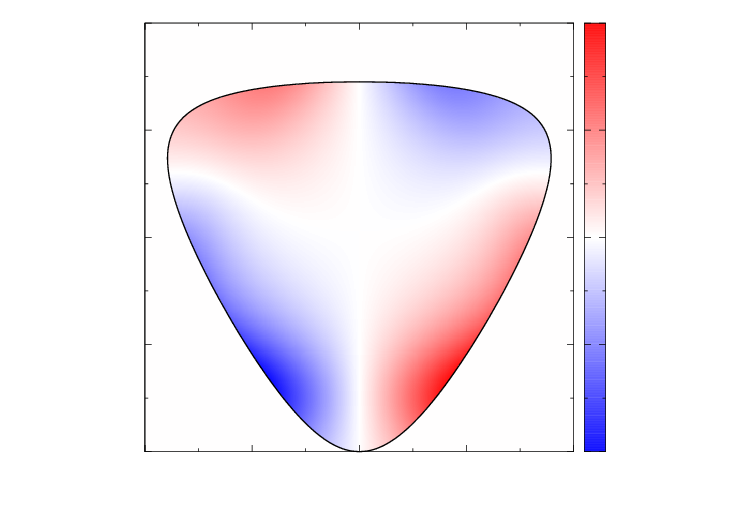}}
    \caption{Estimation for the distribution of the real (left) and imaginary (right) parts of $\M_0^{\not C}$ (top) and $\M_2^{\not C}$ (bottom) in $\eta'\to\pi^+\pi^-\pi^0$ over the allowed phase space. As a rough estimation, the normalizations of each amplitude are fixed by the central fit result obtained for $\eta\to\pi^+\pi^-\pi^0$ in Table~\ref{tab:SubConsEta3Pi}. 
    In contrast to Fig.~\ref{fig:DalitzPlotsEta} additional zero lines occur as a consequence of the interference with the $\rho$-resonance, which lies in the kinematically accessible region.
    }
    \label{fig:DalitzPlotsEtap}
\end{figure}
The thereby generated distributions of the real and imaginary parts of the $C$-odd amplitudes in Fig.~\ref{fig:DalitzPlotsEtap} show that the chiral suppression of the isoscalar transition with respect to the isotensor one is attenuated significantly by the increased phase space, such that $\M_0^{\not C}$ dominates $\M_2^{\not C}$ by roughly two orders of magnitude.  More precisely, we predict that the relative sensitivity to the isoscalar transition is increased by about two orders of magnitude in comparison to the analogous $\eta$ decay.  
This scaling can be qualitatively understood: we have emphasized that the isoscalar $C$-odd operators are suppressed by four orders in the chiral expansion with respect to the isotensor ones.  However, the $\eta'$ decay into three pions is far less a low-energy decay: the available phase space is larger by about a factor of $(M_{\eta'}-3M_\pi)/(M_{\eta}-3M_\pi) \approx 4$.  Taking this to the fourth power correctly predicts an increased relative sensitivity by roughly a factor of 250.

A more rigorous analysis of $C$-odd effects can be performed once a dispersion-theoretical fit to $\eta'\to3\pi$ Dalitz plots within the Standard Model is accomplished~\cite{IskenStoffer:2022}.

\section{Dispersive representation of $\boldsymbol{\eta'\to\eta\pi\pi}$}
\label{sec:ep-epp}
In this section we turn our attention to another class of TOPE forces by studying the decay $\eta'\to\eta\pi^+\pi^-$. Considering the quantum numbers of the involved mesons, one can argue in a similar manner as previously in Sect.~\ref{sec:eta}: as the decay at hand preserves $G$-parity, transitions of even isospin $\Delta I=0,2$ conserve $C$, while odd ones violate the latter. Thus we can write the most general amplitude up to linear order in isospin breaking as 
\beq\label{eq:FullAmpetapEtaPiPi}
    \M(s,t,u)= \M^C_0(s,t,u) + \M^{\not C}_1(s,t,u), 
\eeq
where for this decay, as opposed to $\eta^{(\prime)}\to3\pi$, the isoscalar amplitude $\M^{C}_0$ is isospin- and $C$-conserving, whereas the $\M^{\not C}_1$ violates both quantum numbers. Note that the decay $\eta'\to\eta\pi^+\pi^-$ is sensitive to a different class of $C$- and $CP$-violating operators from those tested in $\etap\to\pi^+\pi^-\pi^0$, namely the ones for transitions with $\Delta I=1$. 

For the evaluation of the overall amplitudes we again rely on the Khuri--Treiman framework, which was already applied to the Standard-Model contribution $\M^C_0$ in Ref.~\cite{Isken:2017dkw}. 
The set of dispersion relations is built from the two scattering processes $\eta'\eta\to\pi\pi$ ($s$-channel) and $\eta'\pi\to\eta\pi$ ($t$-channel). Once more, we allow only for elastic rescattering. In order to determine the $C$-odd amplitude we follow the same agenda as laid out in Sect.~\ref{sec:eta}.

\subsection{Kinematics}
\label{sec:kinematics eta'}
Define the $\eta'\to\eta\pi^+\pi^-$ transition amplitude as usual by
\beq
    \big\langle \pi^+(p_+)\,\pi^-(p_-)\,\eta(p_\eta)\big| iT \big| \eta(P_{\eta'})\big\rangle
    =i\,(2\pi)^4\,\delta^{(4)}( P_{\eta'}-p_+-p_--p_\eta )\,\M(s,t,u)\,.
\eeq
For the invariant masses we stick to the convention
\beq
    s=\left(P_{\eta'}-p_\eta\right)^2\,, \qquad
    t=\left(P_{\eta'}-p_{\pi^+}\right)^2\,, \qquad 
    u=\left(P_{\eta'}-p_{\pi^-}\right)^2\,.
\eeq
These Mandelstam variables satisfy the relation
\beq
    s+t+u=\metap^2+\meta^2+2\mpi^2 \equiv 3r\,.
\eeq
For the $s$-channel scattering amplitude $\eta'\eta\to \pi\pi$
one may write 
\beq
    t(s,z_s) = u(s,-z_s)=\frac{1}{2} \big( 3r-s+ z_s\,\kappapi(s) \big)\,,
\eeq
with
\beq
    z_s\equiv \cos\theta_s=\frac{t-u}{\kappapi(s)}\,,\qquad 
    \kappapi(s)=\sigma(s)\,\lambda^{1/2}(s, \metap^2,\meta^2)\,.
\eeq
For the $t$-channel $\eta'\pi\to \eta\pi$ we have
\beq
    s(t,z_t),\, u(t,z_t)=\frac{1}{2} \Big( 3r-t\mp \frac{\Delta}{t}\mp z_t\,\kappaeta(t) \Big)\,,
\eeq
with $\Delta\equiv( \metap^2-\mpi^2)(\meta^2-\mpi^2)$. Using the kinematic function
\beq
    \kappaeta(t)=\frac{\lambda^{1/2}(t,\metap^2,\mpi^2)\,\lambda^{1/2}(t,\meta^2,\mpi^2)}{t}
\eeq
we can express the $t$-channel scattering angle as
\beq
    z_t\equiv \cos\theta_t=\frac{t\,(u-s)-\Delta}{t\,\kappaeta(t)}\,.
\eeq
As a consequence of crossing symmetry the corresponding relations for the $u$-channel can be obtained by exchanging the variables $t\leftrightarrow u$ and $z_t \leftrightarrow -z_u$. The scattering channels have the physical thresholds
\beq
    \sth=4\mpi^2, \hspace{.5cm} \tth=\uth=\left( \meta+\mpi \right)^2.
\eeq

\subsection{Reconstruction theorem}
\label{sec:reconstruction theorem eta'}

In the ongoing, we restrict our amplitude to discontinuities in the lowest contributing partial waves, i.e., to $\ell=0$ for $\pi\pi$ states with isospin $I=0$, or $\ell=1$ for those with $I=1$, and to $\ell=0$ for the $\eta\pi$ system with $I=1$. We neglect the phase of the $\eta\pi$ $P$-wave, which has exotic quantum numbers (i.e., no resonances are expected in the quark model), and is as suppressed at low energies in the chiral expansion as $D$- and higher partial waves~\cite{Bernard:1991xb}. With these approximations the decomposition of the Standard-Model amplitude in terms of single-variable functions takes the simple form~\cite{Isken:2017dkw}
\begin{equation}\label{eq:reconstruction-theorem-eta'}
	\M^{C}_0(s,t,u) = \F\pipi(s) + \F\etapi(t) + \F\etapi(u)\,, 
\end{equation}
with the abbreviations $\F\pipi(s)\equiv\F^{\ell=0}_{I=0\,\pi\pi}(s)$ and $\F\etapi(t)\equiv\F^{\ell=0}_{I=1\,\eta\pi}(t)$. In this notation the indices $\pi\pi$ and $\eta\pi$ denote the two-particle final state of the respective scattering process. 
In a similar fashion we obtain the reconstruction theorem for the $C$-violating amplitude 
\beq \label{eq:reconstruction-theorem-eta'C}
    \M^{\not C}_1(s,t,u) = (t-u)\,\G\pipi(s) + \G\etapi(t) - \G\etapi(u) \,,
\eeq
which can be read off along the lines of Ref.~\cite{Isken:2021gez}. In this equation we use the short form $\G\pipi(s)\equiv\G{}^1_{1\,\pi\pi}(s)$ and $\G\etapi(s)\equiv\G{}^0_{1\,\eta\pi}(s)$. The ambiguities of these representations are given by the transformations 
\beq
\begin{aligned}
    \F\etapi(t) &\to \F\etapi(t) - \frac{1}{2} a_0 + b_0\,(t-r)\,, & \F\pipi(s) &\to \F\pipi(s) + a_0 + b_0\,(s-r)\,,\\[.1cm]
    \G\etapi(t) &\to \G\etapi(t) + a_1 - b_1\,t +c_1\,t\,(t-3r)\,, & \G\pipi(s) &\to \G\pipi(s) + b_1 + c_1\, s\,, \label{eq:AmbiguitiesEtapEtaPiPi}
\end{aligned}
\eeq
which leave the full amplitudes unaffected.
\subsection{Elastic unitarity}
\label{sec:elastic unitarity eta'}

To ensure the conservation of probability, the single-variable functions have to obey
\beq\label{eq:DiscEtapEtaPiPi}
\begin{aligned}
    \disc \A\pipi(s) &= 2i\,\theta(s-4\mpi^2)\,\big[\A\pipi(s)+\hat \A\pipi(s)\big]\,\sin\delta\pipi(s)\,e^{-i\delta\pipi(s)}\,,\\[.1cm]
    \disc \A\etapi(t) &= 2i\,\theta\big(t-(\meta+\mpi)^2\big)\,\big[\A\etapi(t)+\hat \A\etapi(t)\big]\,\sin\delta\etapi(t)\,e^{-i\delta\etapi(t)}\,,
\end{aligned}
\eeq
with $\A\in\{ \F, \G \}$ and the indices of the phase shifts labeling the respective two-particle intermediate states. Note that in case of $\M_0$ the $\pi\pi$-state has isospin $I=0$, such that $\delta\pipi=\delta^{I=0}\pipi$ for $\A=\F$. Analogously, the $C$-odd contribution $\M_1^{\not C}$ is driven by a $\pi\pi$-state that has isospin $I=1$, i.e., $\delta\pipi=\delta^{I=1}\pipi$ for $\A=\G$.
Introducing the abbreviations
\beq
\begin{aligned}
 &\hspace{3.5cm}\langle z_{s}^n\,\A\rangle \equiv\frac{1}{2}\int_{-1}^{1}\diff z_{s}\,z_{s}^n\, \A\big(t(s,z_{s})\big)\,,\\[.1cm]
 &\langle z_{t}^n\,\A\rangle^{+}\equiv\frac{1}{2}\int_{-1}^{1}\diff z_{t}\,z_{t}^n\,\A\big(u(t,z_{t})\big)\,,\qquad
 \langle z_{t}^n\,\A\rangle^{-} \equiv\frac{1}{2}\int_{-1}^{1}\diff z_{t}\,z_{t}^n\,\A\big(s(t,-z_{t})\big)\,,
 \end{aligned}
\eeq
the inhomogeneities for the Standard-Model amplitude, obtained by a partial-wave projection as described in Sect.~\ref{sec:elastic unitarity eta}, become 
\begin{equation}
\hat \F\pipi(s) = 2\langle\F\etapi\rangle\,, \qquad
\hat \F\etapi(t) = \langle\F\pipi\rangle^- + \langle\F\etapi\rangle^+ \,,
\end{equation}
and the ones entering the $C$-violating amplitude yield
\beq
\begin{aligned}
    \hat\G\pipi(s)&=\frac{6}{\kappa\pipi}\,\langle z_{s}\,\G\etapi \rangle\,,\\[.1cm]
    \hat\G\etapi(t)&=-\langle \G\etapi \rangle^+ -\frac{3}{2}\bigg( r-t+\frac{\Delta}{3t}\bigg)\,\langle  \G\pipi \rangle^- + \frac{1}{2}\kappa\etapi\,\langle z_{t}\,\G\pipi \rangle^-\,.
\end{aligned}
\eeq
Analogously to Eq.~\eqref{eq:DispersionIntegral} we can write the general solutions as
\beq\label{eq:DispIntEtapEtaPiPi}
\begin{aligned}
    \A\pipi(s)&= \Omega\pipi(s)\,\bigg(P^{n-1}\pipi(s)+\frac{s^n}{\pi}\int_{\sth}^\infty \frac{\diff x}{x^{n}}\, \frac{\sin\delta\pipi(x)\,\hat{\A}\pipi(x)}{|\Omega\pipi(x)|\, (x-s)}\bigg)\,,\\[.1cm]
    \A\etapi(t)&= \Omega\etapi(t)\,\bigg(P^{n-1}\etapi(t)+\frac{t^n}{\pi}\int_{\tth}^\infty \frac{\diff x}{x^{n}}\, \frac{\sin\delta\etapi(x)\,\hat{\A}\etapi(x)}{|\Omega\etapi(x)|\, (x-t)}\bigg)\,,
\end{aligned}
\eeq
with two distinct subtraction polynomials $P^{n-1}\pipi$ and $P^{n-1}\etapi$ of order $n-1$. The index of each Omnès function decides which scattering phase shift is used according to Eq.~\eqref{eq:Omnes}. In addition to that, one has to differentiate the case $\Omega\pipi=\Omega^{I=0}\pipi$ for $\A=\F$ from $\Omega\pipi=\Omega^{I=1}\pipi$ for $\A=\G$.

\begin{figure}[t!]
    \centering
	\hspace{-.75cm}\scalebox{.65}{\input{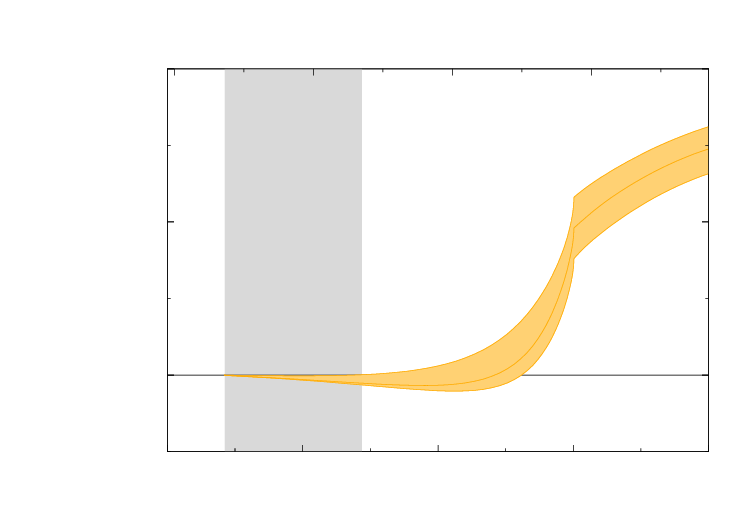}}\hspace{-.75cm}\hfill
	\scalebox{.65}{\input{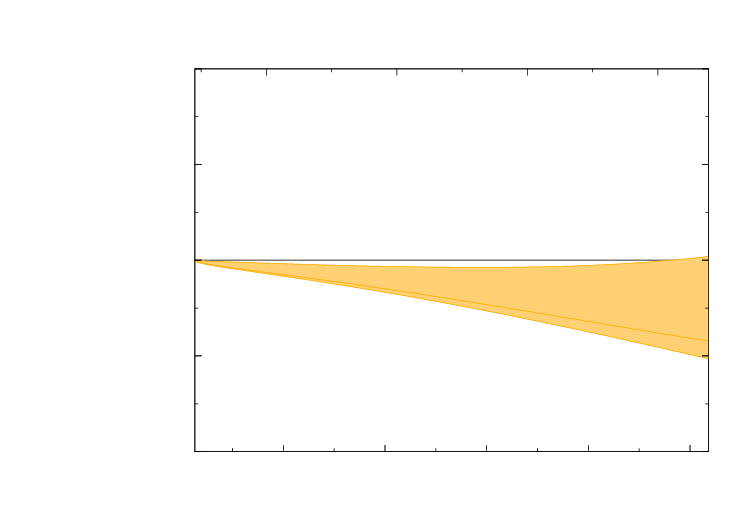}}
	\caption{$S$-wave $\eta\pi$ phase shift extracted from Refs.~\cite{Albaladejo:2015aca,Lu:2020qeo} including its uncertainty band. Left panel: behavior of the phase shift in the low- and intermediate-energy region. The $K\bar{K}$-cusp is clearly visible at about $50\mpi^{2}$. The phase space for $\eta'\to\eta\pi\pi$ is indicated by the gray region. Right panel: magnification of the physical decay region.
	}
	\label{fig:Phase_PiEta}
\end{figure}
As our numerical input, we use the same $\pi\pi$ phase shifts as detailed in the discussion of $\eta\to3\pi$ in Sect.~\ref{sec:SubSchemeEta}.  For the $\eta\pi$ $S$-wave, we employ the phase of the corresponding scalar form factor constructed in Ref.~\cite{Albaladejo:2015aca}, further refined by imposing constraints from $\gamma\gamma\to\eta\pi^0$~\cite{Lu:2020qeo}.  This phase, including the associated uncertainties, is shown in Fig.~\ref{fig:Phase_PiEta}.

\subsection{Subtraction scheme}
\label{sec:SubScheme Eta'}
In this section we proceed in the same fashion as in Sect.~\ref{sec:SubSchemeEta} to fix the yet undetermined number of subtractions entering the dispersive representation in Eq.~\eqref{eq:DispIntEtapEtaPiPi}. 
We assume that the involved phase shifts behave in the limits $s\to\infty$ or $t\to\infty$, respectively, as
\beq
    \delta^0\pipi(s)\to \pi\,, \qquad \delta^1\pipi(s)\to \pi\,, \qquad \delta\etapi(t)\to \pi\,.
\eeq
Furthermore, we demand the asymptotics of the single-variable functions $\F\pipi$ and $\F\etapi$ resulting from the Froissart--Martin bound~\cite{Froissart:1961ux}
\beq
    \F\pipi(s)=\mathcal{O}(s)\,, \qquad \F\etapi(t)=\mathcal{O}(t)\,.
\eeq
This results in a representation of the corresponding SVAs involving four (real) subtraction constants, 
\beq\label{eq:SM-SVAs eta'}
\begin{aligned}
			\F\pipi(s)&= \Omega\pipi^0(s)\,\bigg(\alpha+\beta\,s+\gamma\,s^2+\frac{s^3}{\pi}\int_{\sth}^\infty \frac{\diff x}{x^{3}}\, \frac{\sin\delta\pipi^0(x)\,\hat{\F}\pipi(x)}{|\Omega\pipi^0(s')|\,(x-s)}\bigg)\,,\\[.1cm]
			\F\etapi(t)&= \Omega\etapi(t)\,\bigg(\lambda\, t^2+\frac{t^3}{\pi}\int_{\tth}^\infty \frac{\diff x}{x^{3}} \frac{\sin\delta\etapi(x)\,\hat{\F}\etapi(x)}{|\Omega\etapi(x)|\, (x-t)}\bigg)\,.
\end{aligned}
\eeq
In Ref.~\cite{Isken:2017dkw}, a more rigorous scheme with asymptotics analogous to those discussed for $\eta\to3\pi$ in Sect.~\ref{sec:SubSchemeEta} and, correspondingly, less subtractions was employed in parallel, and found to describe the Dalitz plot data similarly well, while being more susceptible to sizeable uncertainties due to high-energy input to the dispersion integrals.  With the adjusted input for $\eta\pi$ scattering~\cite{Lu:2020qeo}, this reduced scheme ceases to work well~\cite{IskenStoffer:2022}.  We regard this partly as an artifact of the extremely slow asymptotic rise of the $\eta\pi$ phase shift, cf.\ Fig.~\ref{fig:Phase_PiEta}, and therefore decide to stick to the more restrictive asymptotics for the $C$-odd contribution all the same, in order to avoid a proliferation of subtraction constants therein.  The assumptions for $\G\pipi$ and $\G\etapi$ hence are 
\beq
     \G\pipi(s)=\mathcal{O}(s^{-1}) \,, \qquad \G\etapi(t)=\mathcal{O}(t^0) \,,
\eeq
such that the resulting $C$-violating SVAs are given by
\beq\label{eq:BSM-SVAs eta'}
\begin{aligned}
			\G\pipi(s)&= \Omega\pipi^1(s)\,\bigg(\varrho +\frac{s}{\pi}\int_{\sth}^\infty \frac{\diff x}{x}\, \frac{\sin\delta\pipi^1(x)\,\hat{\G}\pipi(x)}{|\Omega\pipi^1(x)|\,(x-s)}\bigg)\,,\\[.1cm]
			\G\etapi(t)&= \Omega\etapi(t)\,\bigg(\zeta\, t+\frac{t^2}{\pi}\int_{\tth}^\infty \frac{\diff x}{x^{2}} \frac{\sin\delta\etapi(x)\,\hat{\G}\etapi(x)}{|\Omega\etapi(x)|\,(x-t)}\bigg)\,.
\end{aligned}
\eeq
Conventionally, the polynomial ambiguities from Eq.~\eqref{eq:AmbiguitiesEtapEtaPiPi} were shifted such that a minimal number of subtraction constants contributes to the $\A\etapi$.
Again, the phase of the subtraction constants $\varrho$ and $\zeta$ is fixed by $T$ violation, so that $\M_1^{\not C}$ has two \textit{real}-valued degrees of freedom, in contrast to the $C$-violating isoscalar and isotensor contributions in $\eta\to3\pi$ which are fixed by a single normalization each.
The numerical implementation proceeds in analogy to the strategy presented in Sect.~\ref{sec:SubSchemeEta}.
\subsection{Taylor invariants}
\label{sec:TaylorInvariantsEta'}
As pointed out in Sect.~\ref{sec:TaylorInvariants Eta}, the subtraction constants fixing our dispersive representation are no suitable observables. Therefore we again introduce their linear combinations as ambiguity-free Taylor invariants obtained by an expansion of the SVAs around $s,t=0$, i.e., 
\beq
\begin{split}
    \A\pipi(s)&=A\pipi^{\A}+B\pipi^{\A}\,s+C\pipi^{\A}\,s^2+D\pipi^{\A}\,s^3+\ldots\,,
    \\[.1cm]
    \A\etapi(t)&=A\etapi^{\A}+B\etapi^{\A}\,t+C\etapi^{\A}\,t^2+D\etapi^{\A}\,t^3+\ldots\,.
\end{split}
\eeq
Of course the series coefficients take different values for SM and BSM contributions. Applying these expansions to the reconstruction theorem~\eqref{eq:reconstruction-theorem-eta'} allows us to express the SM amplitude by
\beq
\M_0^C(s,t,u)=F_0+F_1\,(2s-t-u)+F_2\,s^2+F_3\,(t^2+u^2)+\mathcal{O}(p^6)
\eeq
with 
\beq
    F_0=A_{\pi\pi}^{\F}+r\,B_{\pi\pi}^{\F}+2(A_{\eta\pi}^{\F}+r\,B_{\eta\pi}^{\F})\,, \quad F_1=\frac{1}{3}\left(B_{\pi\pi}^{\F}-B_{\eta\pi}^{\F}\right)\,, \quad F_2=C_{\pi\pi}^{\F}\,, \quad F_3=C_{\eta\pi}^{\F}\,,
\eeq
where we dropped terms of cubic order in the Mandelstam variables and higher. The BSM operator driving the $\Delta I=1$ transition as introduced in Eq.~\eqref{eq:BSM-operator etap} demands that the matrix element takes the form
\beq\label{eq:BSMCouplingI=1}
    \M_1^{\not C}(s,t,u) = i\,g_1\,(t-u) \left(1+s\,\delta g_1\right)+\Order(p^6)  \,,
\eeq
where in addition to the effective isovector coupling $g_1$, we also consider the leading $s$-dependent correction $\delta g_1$. In terms of the Taylor coefficients these quantities read
\beq\label{eq:g1coupling}
    g_1= -i\,\big(A\pipi^{\G}+B\etapi^{\G}+3r\,C\etapi^{\G}\big)\,,\qquad \delta g_1=-i\,\big( B\pipi^{\G}-C\etapi^{\G}\big)/g_1\,.
\eeq
Note that the additional parameter $\delta g_1$ ensures that the degrees of freedom of the Taylor expansion match the ones of the dispersive representation for $\M_1^{\not C}$.
Both couplings are real-valued as demanded by $T$ violation and give rise to the phases of the subtraction constants $\varrho$ and $\zeta$. The latter can be considered as purely imaginary due to the small available phase space.

\subsection{Fixing the subtraction constants}
\label{sec:FixingSubConsEta'}
According to the subtraction scheme chosen in Sect.~\ref{sec:SubScheme Eta'}, the dispersive representation of the SM amplitude $\M_0^C$ contains the four degrees of freedom $\alpha$, $\beta$, $\gamma$, $\lambda$, where again one subtraction constant can be chosen to fix the overall normalization. The $C$-violating isovector contribution $\M_1^{\not C}$ has a total of three parameters $\varrho$, $\zeta$, and $\varphi$, where the latter fixes the complex phase between $\M_0^C$ and $\M_1^{\not C}$. After solving for the basis solutions of the dispersive representation in Eqs.~\eqref{eq:SM-SVAs eta'} and \eqref{eq:BSM-SVAs eta'}, these subtraction constants can be determined by a comparison to data.

In contrast to Sect.~\ref{sec:FixingSubcons Eta}, we only consider one single data set, i.e., the Dalitz-plot distribution $\mathcal{D}$ of $\eta'\to\eta\pi^+\pi^-$ from the BESIII collaboration~\cite{BESIII:2017djm}. The latter provides the currently most precise measurement including $3.51\times10^{5}$ events extracted from $J/\psi$ decays in terms of the symmetrized coordinates
\begin{align}\label{eq:XYdefEtapEtaPiPi}
x &= 
\frac{\sqrt{3}}{2 M_{\eta'} Q_{\eta'}} (u-t)\,,\qquad
y = \frac{( M_\eta+2 M_\pi)}{2 M_\pi M_{\eta'} Q_{\eta'}}\big[(M_{\eta'}-M_{\eta})^2-s\big]-1\,, 
\end{align}
with $Q_{\eta'} = M_{\eta'}-M_\eta - 2 M_{\pi}$. 
We refrain from including data sets on $\eta'\to\eta\pi^0\pi^0$~\cite{Blik:2009zz,Adlarson:2017wlz,BESIII:2017djm} in the analysis, as, in contrast to the case of $\eta\to3\pi^0$, they do not provide truly independent information on the SM amplitude, but rather probe subtle isospin-breaking effects~\cite{Kubis:2009sb,Isken:2017dkw}.
We determine the subtraction constants by minimizing the discrepancy function
\begin{equation}
	\chi^{2} = \sum_{i}\bigg(\frac{\mathcal{D}(x_{i},y_{i})-|\M(x_{i},y_{i})|^2}{\Delta \mathcal{D}(x_{i},y_{i})}\bigg)^{2}\,,
\end{equation}
for which we compute our dispersive amplitude $\M$ on the discrete grid covering the centers of all measured bins and normalize $\M$ to reproduce the according experimental decay width $\Gamma(\eta'\to\eta\pi^+\pi^-)=79.9(2.7)\,\text{keV}$ taken from the PDG~\cite{Zyla:2020zbs}.

\begin{figure}
    \centering
	\scalebox{.6}{\input{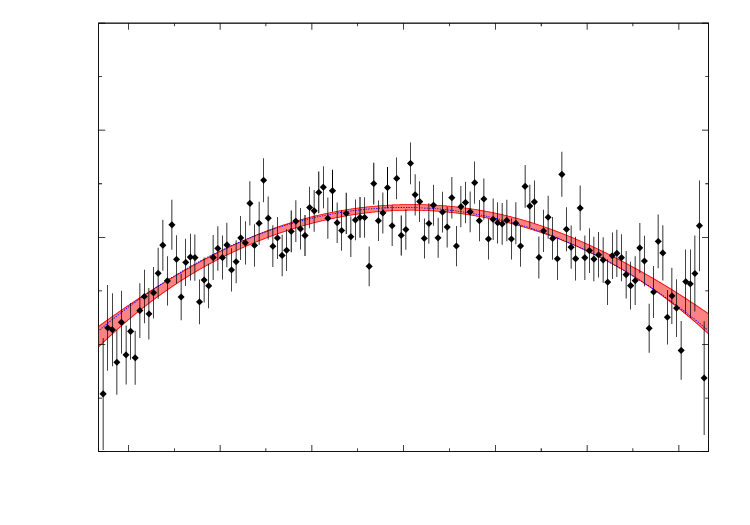}}\hspace{-.75cm}\hfill
	\scalebox{.6}{\input{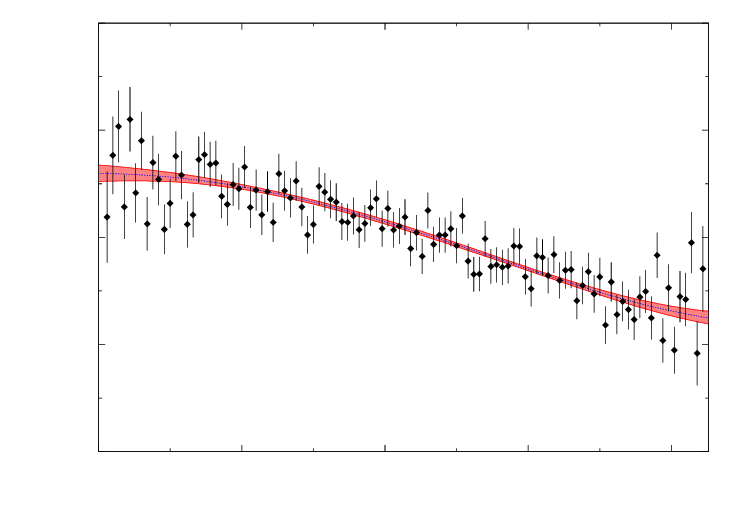}}
    \caption{Dalitz-plot projections in $x$- and $y$-direction, which are divided by the corresponding phase space. We show the measurement of Ref.~\cite{BESIII:2017djm} overlayed with our dispersive representations covered by the red error bands. Note that the theoretical $x$-projection on the left is not perfectly symmetric, due to $C$-violating contributions. In both panels we depict our central solution for the $C$-conserving part $|\M_{0}^{C}|^{2}$ by the dotted blue line.}
    \label{fig:ProjectionsEtapEtaPiPi}
\end{figure}

We proceed by carrying out the regression using the pure SM amplitude $\M_0^C$ as well as the one for the full BSM contribution $\M = \M_0^C + \M_1^{\not C}$.  The results for these fit scenarios, denoted as $\text{FIT}_\text{SM}$ and $\text{FIT}_\text{BSM}$, are listed in Table~\ref{tab:Fit Eta'} and the corresponding subtraction constants can be found in Table~\ref{tab:SubConsEta'}. 
\begin{table}[t]
\centering
\renewcommand{\arraystretch}{1.5}
\begin{tabular}{lcccc}
\toprule
 & $\chi^{2}$ & dof & ~$\chi^{2}/\text{dof}$~ & ~$p$-value\\
\midrule
$\text{FIT}_\text{SM}$~~ & ~ 10720 ~ &  ~ 10790 ~ &  0.994  & ~$ 68\%$ \\
$\text{FIT}_\text{BSM}$~~  & ~ 10718 ~ & ~ 10788 ~ & 0.994 & ~$ 68\%$ \\
\bottomrule
\end{tabular}
\renewcommand{\arraystretch}{1.0}
\caption{Goodness of the central fit results for the SM amplitude $\M_0^C$ (FIT$_\text{SM}$) and the full one $\M_1^{\not C}$ (FIT$_\text{BSM}$) obtained by comparison with the BESIII data set~\cite{BESIII:2017djm}.
}
\label{tab:Fit Eta'}
\end{table}
We observe that the additional inclusion of the $C$-violating $\Delta I=1$ transition does not have any visible influence on the overall goodness of the regression. As an illustration of the latter we show the phase space corrected $x$- and $y$-projections of the Dalitz plot in Fig.~\ref{fig:ProjectionsEtapEtaPiPi}. Note that the small effects of mirror symmetry breaking are apparent in the $x$-projection.
\begin{table}[t]
	\centering
	\renewcommand{\arraystretch}{1.5}
	\setlength{\tabcolsep}{5pt}
	\setlength\extrarowheight{2pt}
	\begin{tabular}{lcccccc}
		\toprule
		& $\alpha$ & $\beta\cdot M_\pi^{2}$ & $\gamma\cdot M_\pi^{4}$ & $\lambda\cdot M_\pi^{4}$& $\Im\varrho\cdot M_\pi^{2}$ & $\Im\zeta\cdot M_\pi^{2}$\\
 		\midrule
	$\text{FIT}_\text{SM}$~~ &$-19.0(8)$ & $1.27(7)$ & $0.0016(30)$ & $0.0060(3)$ &-- & -- \\
	$\text{FIT}_\text{BSM}$~~ &$-19.0(8)$ & $1.27(7)$ & $0.0016(30)$ & $0.0060(3)$ & $-0.04(12)$ &$0.05(12)$  \\
	\bottomrule
	\end{tabular}
\renewcommand{\arraystretch}{1.0}
	\caption{Results for the subtraction constants of the Standard-Model amplitude in the first row and the full $C$- and $CP$-odd dispersive representation in the second row.}
	\label{tab:SubConsEta'}
\end{table}

Due to the fact that the current constraints for the $\eta'\to\eta\pi^+\pi^-$ SM amplitude are by far less restrictive than the ones pointed out for $\eta\to3\pi$ in Sect.~\ref{sec:SM constraints eta}, we omit an elaborate analysis of the asymmetric systematical errors when varying the input for the $\eta\pi$ phase shift shown in Fig.~\ref{fig:Phase_PiEta}~\cite{IskenStoffer:2022}. However, we remark that these systematical errors for the SM amplitude may increase up to the same order of magnitude as the corresponding statistical ones. In either way, the $C$-violating observables in the central scope of this analysis are dominated by their statistical uncertainties.

\subsection{Extraction of observables}
\label{sec:observables eta'}

In this section we work out the numerical results of our dispersive representation for various $C$-violating observables in the $\eta'\to\eta\pi^+\pi^-$ amplitudes. Similar to Sect.~\ref{sec:Observables Eta} we first discuss the validity of our SM amplitude. To this end we extract the Adler zeros and the Taylor invariants in Sect.~\ref{sec:SM constraints eta'}. Thereafter we extract patterns of $C$-violation in the Dalitz-plot distribution, investigate the occurring asymmetries, and compute the coupling strength of an effective isovector BSM operator $X_1^{\not C}$. 

\subsubsection{Standard Model constraints}
\label{sec:SM constraints eta'}
\begin{sloppypar}

The Taylor invariants $F_i$ defined in Sect.~\ref{sec:TaylorInvariantsEta'} allow us to extract coefficients that can be compared to theoretical analyses for the $\eta'\to\eta\pi^+\pi^-$ SM contribution as for instance large-$N_c$ $\chi$PT or R$\chi$T~\cite{Escribano:2010wt,Gonzalez-Solis:2018xnw}. 

As described in Sect.~\ref{sec:SubScheme Eta'}, we use four real-valued subtraction constants to fix the SM amplitude. These can be translated to the Taylor invariants
\beq
	\begin{array}{rr|cccc}
		F_0\phantom{/\text{GeV}^{-2}} = & -13.0(7)\, & 1.00 & -0.67 & \phantom{+}0.91 & -0.49 \\[.1cm]
		f_1/\text{GeV}^{-2} = &\ -0.3(1)\, & \phantom{1.00} & 1.00 & -0.86 & \phantom{+}0.97 \\[.1cm]
		f_2/\text{GeV}^{-4} = &\! \phantom{+}3.0(4)\, & \phantom{1.00} & \phantom{+1.00} & \phantom{+}1.00 & -0.72 \\[.1cm]
		f_3/\text{GeV}^{-4} = &\ -1.2(1)\, & \phantom{1.00} & \phantom{1.00}  & \phantom{1.00} & \phantom{+} 1.00
\end{array}\,,
\label{eq:SM Taylor Coeffs Eta'}
\eeq
where $F_{0}$ serves as an overall normalization by means of $f_{i}\equiv F_{i}/F_{0}$. Possible imaginary parts of the Taylor invariants are exclusively generated by the dispersion integrals \eqref{eq:SM-SVAs eta'} and are disregarded in the following.

Furthermore we want to study the behavior of the SM amplitude outside the physical region at its soft-pion points. Chiral SU(2)$_R\times$SU(2)$_L$ symmetry expects two Adler zeros to show up at $(t-u)=\pm(\metap^2-\meta^2)$ along the line $s=0$ in the limit of massless pions~\cite{Adler:1965a,Adler:1965b,Riazuddin:1971}. Therefore, in analogy to Ref.~\cite{Isken:2017dkw} we study our dispersive amplitude for on-shell pions along the critical line $s=2\mpi^{2}$ and find two zeros at
\beq
    (t_{A}-u_{A})/(\metap^2-\meta^2) = \pm0.902(23)\,.
\eeq

An updated analysis of the SM $\eta'\to\eta\pi\pi$ decay presented in Ref.~\cite{Isken:2017dkw} is currently in progress~\cite{IskenStoffer:2022}, based on the latest high-statistics Dalitz-plot measurements from A2~\cite{Adlarson:2017wlz} and BESIII~\cite{BESIII:2017djm} for the charged and neutral decay modes.  

\end{sloppypar}

\subsubsection{Dalitz-plot distribution}
\label{sec:Dalitzplots eta'}
\begin{figure}[t!]
    \centering
    \hspace{-.25cm}\scalebox{.63}{\input{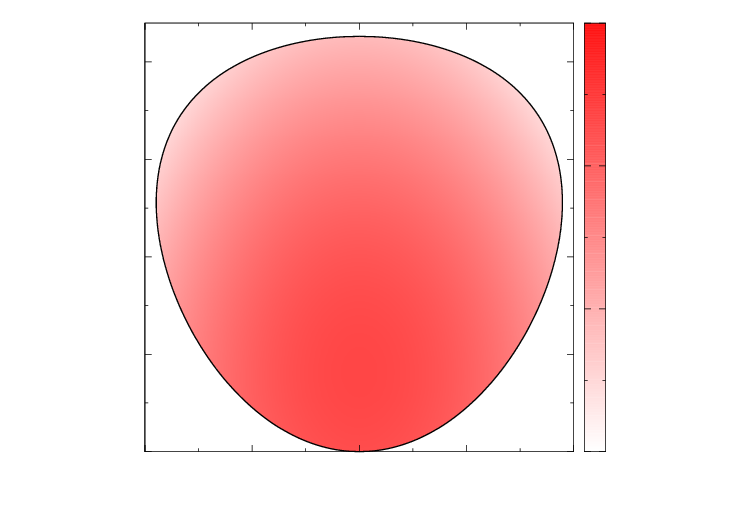}}\hspace{-.75cm}\hfill
    \scalebox{.63}{\input{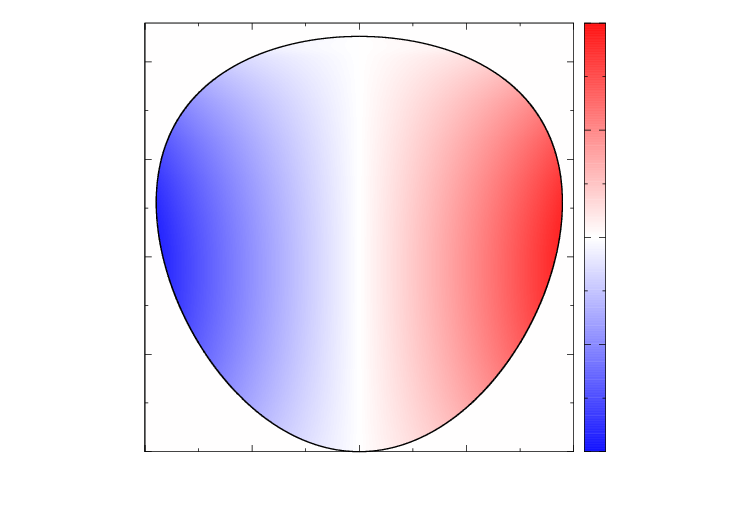}}
    \caption{Dalitz-plot decomposition for $\eta'\to\eta\pi^+\pi^-$ as given in Eq.~\eqref{eq:DalitzDecompEtapEtaPiPi} for our central solution. The normalization is chosen such that the full amplitude $|\M|^2$ is one in its center. The interference term of $\M_0^C$ and $\M_1^{\not C}$ gives rise to the breaking of mirror symmetry. Note the relative order of magnitudes between the individual contributions.}
    \label{fig:DalitzPlotsEtapEtaPiPi}
\end{figure}

Let us continue our discussion on $C$-violating patterns arising from the  $\Delta I=1$ transition $\eta'\to\eta\pi^+\pi^-$ Dalitz-plot distribution and quantify corresponding observables. Dropping the dependencies on the coordinates $x$ and $y$ and neglecting the contribution of $|\M_1^{\not C}|^2$, the Dalitz-plot distribution arising from Eq.~\eqref{eq:FullAmpetapEtaPiPi} can be written as
\beq\label{eq:DalitzDecompEtapEtaPiPi}
    |\M|^2\approx|\M_0^{C}|^2 + 2\,\mathrm{Re}\left[\M_0^{C}\, (\M_1^{\not C})^\ast\right]\,,
\eeq
which is depicted in Fig.~\ref{fig:DalitzPlotsEtapEtaPiPi}. 
We observe a similar, however slightly flattened, hierarchy as in the case of $\eta\to3\pi$ worked out in Sect.~\ref{sec:Dalitzplots Eta 3 pi}. The interference term giving rise to the Dalitz-plot asymmetry is constrained to be  three orders of magnitude smaller than the SM contribution $|\M_0^{C}|^2$, whereas the pure $\Delta I=1$ contribution $|\M_1^{\not C}|^2$ is suppressed by four orders of magnitude. We conclude that the current state of precision for the $\eta'\to\eta\pi^+\pi^-$ Dalitz plot merely restricts the effects of the $C$-violating isovector transition to the relative per mille level. 

Given the small phase space of the process, the momentum distribution is quite smooth and commonly approximated by the same expansion as introduced in Eq.~\eqref{eq:PolynomialExpansion}, but with adapted coordinates $x$ and $y$ from Eq.~\eqref{eq:XYdefEtapEtaPiPi}. The BESIII collaboration finds that the first three $C$-even coefficients $a$, $b$, and $d$ of this expansion are sufficient to parameterize the Dalitz plot, as all other parameters of higher orders in $x$ and $y$, as well as all parameters odd in $x$ indicating $C$-violation, are found to be compatible with zero within less than one standard deviation.
A two dimensional Taylor expansion around the center of our dispersive representation of the Dalitz plot gives rise to the parameters
\begin{equation}
	\begin{array}{rr|ccccc}
		a = & -0.058(4)\, & 1.00 & -0.32 & -0.01 & -0.21 & -0.02\\[.1cm]
		b = & -0.050(7)\, & \phantom{1.00} & \phantom{+}1.00 & \phantom{+}0.00 & \phantom{+}0.32 & -0.01\\[.1cm]
		c = &\ 0.004(3)\, & \phantom{1.00} & \phantom{+1.00} & \phantom{+}1.00 & \phantom{+}0.00 & -0.16\\[.1cm]
		d = & -0.063(4)\, & \phantom{1.00} & \phantom{+1.00} &\phantom{+1.00} & \phantom{+}1.00 & -0.02\\[.1cm]
		e = &\ 0.000(7)\, & \phantom{1.00} & \phantom{+1.00} & \phantom{+1.00} &\phantom{+1.00} & \phantom{+}1.00
\end{array}\,.
\label{eq:dalc-params-cc_eta'}
\end{equation}
where we neglect correlations smaller than 1$\%$ on the right-hand side. 
Considering the respective errors we find a perfect agreement of all our parameters with the experiment~\cite{BESIII:2017djm}. In particular there is no indication for $C$-violation as $c$ and $e$ are effectively zero.

\subsubsection{Asymmetry and BSM coupling}
\label{sec:asymmetries eta'}
To finalize our analysis we quantify the asymmetry and the coupling strength of the $\Delta I=1$ transition in $\eta'\to\eta\pi^+\pi^-$ and apply the same procedure as in Sect.~\ref{sec:AsymmEta3Pi}. 
We find the left-right asymmetry in units of $10^{-3}$ to be\begin{equation}
		A_{LR} = 2.1(1.5).
\label{eq:Coupling I=1}
\end{equation}
Thus the mirror symmetry breaking vanishes within roughly $1.4\sigma$. Furthermore, we can parameterize $A_{LR}$ in terms of the Taylor invariants
\begin{equation}
\begin{array}{rr|ccc}
	g_{1}/\text{GeV}^{-2} = & 0.7(1.0)\, & 1.00 & -0.89 \\[.1cm] 	\delta g_{1}/\text{GeV}^{-2} =& -5.5(7.3) & \phantom{1.00} & \phantom{+}1.00
\end{array}\,.
\label{eq:BSM-operators-coupling-strengths-Etap}
\end{equation}
which were introduced in Eq.~\eqref{eq:g1coupling} as the effective isovector coupling $g_{1}$ and its leading $s$-dependent correction $\delta g_{1}$, respectively. This allows us to write the left-right asymmetry, again in units of $10^{-3}$, in the compact form
\beq
    A_{LR} = 6.6\,g_1\big(1+0.10\,\delta g_1\big)\,,
\eeq
where $g_1$ and $\delta g_1$ enter in units of $\text{GeV}^{-2}$.

\section{Summary}
\label{sec:summary}

In this study, we have put the pioneering work of Ref.~\cite{Gardner:2019nid} for $C$- and $CP$-violating amplitude representations in the decay $\eta\to\pi^+\pi^-\pi^0$ into a rigorous dispersion theoretical framework, and extended the formalism to the analysis of $C$- and $CP$-violation in the hadronic three-body decays of the $\eta'$. 
Strictly relying on the fundamental principles of analyticity and unitarity, we constructed all three $\eta\to\pi^+\pi^-\pi^0$ amplitudes of distinct total isospin, i.e., the SM amplitude $\M_1^C$ as well as the $C$-violating isoscalar and isotensor contributions $\M_0^{\not C}$ and $\M_2^{\not C}$, non-perturbatively based on $\pi\pi$ phase shifts.  We demonstrated that the same constraints---all amplitudes are not allowed to grow asymptotically for large energies---allow us to describe the experimental data by the KLOE-2 collaboration~\cite{Anastasi:2016cdz}, fulfill constraints from chiral perturbation theory on $\M_1^C$, and reduce the freedom in the $C$-violating amplitudes to only one single complex normalization constant each. The phase of the latter is fixed by hermiticity and $T$ violation, resulting in one real-valued free parameter for the isoscalar and isotensor transition, respectively.
Ensuring that the Standard-Model contribution is in good accordance with the dispersive representation of Ref.~\cite{Colangelo:2018jxw}, we extracted the contributions of $\M_0^{\not C}$ and $\M_2^{\not C}$, whose interference with $\M_1^C$ give rise to the breaking of mirror symmetry in the $\eta\to\pi^+\pi^-\pi^0$ Dalitz-plot distribution. We confirmed that the currently most precise measurement of the latter restricts the $C$-violating effects to a relative per mille level.  Due to the strong kinematic suppression of $\M_0^{\not C}$---the corresponding operator is smaller by four orders in the chiral expansion compared to $\M_2^{\not C}$---the accompanying effective coupling constant $g_0$ is far less rigorously constrained than $g_2$, by about three orders of magnitude.

Although there is no sufficiently precise Dalitz-plot measurement for $\eta'\to\pi^+\pi^-\pi^0$ yet, we have demonstrated that, in principle, the larger available phase space lifts the suppression of the isoscalar $C$-odd amplitude to a large extent, making a potential mirror-symmetry breaking therein more sensitive to $\M_0^{\not C}$ by roughly two orders of magnitude than in $\eta\to\pi^+\pi^-\pi^0$.  Both decays would most likely be driven by the same, fundamental, BSM operators.

In a similar manner, we established a framework to analyze the decay $\eta'\to\eta\pi^+\pi^-$, which is sensitive to another class of $C$- and $CP$-violating operators with isospin $I=1$. In this decay the amplitude decomposes into the isoscalar SM amplitude $\M_0^C$ and a $C$-violating isovector contribution $\M_1^{\not C}$. A regression to the Dalitz plot of the BESIII collaboration~\cite{BESIII:2017djm} yields again no evidence for $C$-violating effects and limits their patterns to a relative per mille level.

The extracted coupling strengths of the underlying effective isoscalar and isotensor BSM operators from $\eta\to\pi^+\pi^-\pi^0$ and the one of the isovector BSM operator from $\eta'\to\eta\pi^+\pi^-$ may in the future be matched to fundamental BSM operators on the quark level, thus allowing to predict a corresponding scale for BSM physics~\cite{Khriplovich:1990ef,Conti:1992xn,Engel:1995vv,Ramsey-Musolf:1999cub,Kurylov:2000ub,Shi:2017ffh}.
We conclude that our framework opens a window to the systematic analysis of $C$- and $CP$-violation in $\eta^{(\prime)}\to\pi^{+}\pi^{-}\pi^{0}$ and $\eta'\to\eta\pi^+\pi^-$ Dalitz plots provided by future high-statistics experimental measurements~\cite{Gan:2015nyc,Gatto:2016rae,Gan:2017kfr,Gatto:2019dhj,Beacham:2019nyx}.

\acknowledgments
We thank Susan Gardner, Martin Hoferichter, and Peter Stoffer for numerous, most useful discussions.  We are grateful to Andrzej Kup\'s\'c for help with the data from Ref.~\cite{BESIII:2017djm}, and to Bachir Moussallam for providing us with the $\eta\pi$ phase shift parameterizations of Ref.~\cite{Lu:2020qeo}. H.A.\ thanks Malwin Niehus for his assistance in developing the numerical algorithm for solving the Khuri--Treiman equations.
Financial support was provided by
the DFG (CRC 110, ``Symmetries and the Emergence of Structure in QCD'')
and the Avicenna-Studienwerk e.V.\ with funds from the BMBF.

\bibliographystyle{JHEP_mod}
\bibliography{base}

\end{document}